\documentclass[12pt]{article}
\usepackage{latexsym,amsmath,amsthm,amssymb,amscd,bbm,graphicx}
\usepackage[mathscr]{euscript}

\usepackage{bigints}

\usepackage{algorithm}
\usepackage{algpseudocode}
\usepackage{float}
\newfloat{algorithm}{t}{lop}
\usepackage{caption}
\usepackage{array,multirow}

\usepackage[title]{appendix}

\usepackage[english]{babel}
\usepackage[nottoc]{tocbibind}
\usepackage{authblk}

\usepackage[pdftex]{hyperref}
\usepackage{color}
\usepackage{fullpage}
\usepackage{setspace}
\date{}
\title{Review of Clustering Methods for Functional Data}
\usepackage{authblk}
\author{\normalsize{Mimi Zhang$^{1, 3}$, Andrew Parnell$^{2, 3}$}}
\affil{\small{$^1$School of Computer Science and Statistics, Trinity College Dublin, Ireland\\
$^2$Hamilton Institute, Maynooth University, Ireland\\
$^3$I-Form Advanced Manufacturing Research Centre, Science Foundation Ireland, Ireland}}

\begin{document}
\maketitle
\begin{abstract}
Functional data clustering is to identify heterogeneous morphological patterns in the continuous functions underlying the discrete measurements/observations. Application of functional data clustering has appeared in many publications across various fields of sciences, including but not limited to biology, (bio)chemistry, engineering, environmental science, medical science, psychology, social science, etc. The phenomenal growth of the application of functional data clustering indicates the urgent need for a systematic approach to develop efficient clustering methods and scalable algorithmic implementations. On the other hand, there is abundant literature on the cluster analysis of time series, trajectory data, spatio-temporal data, etc., which are all related to functional data. Therefore, an overarching structure of existing functional data clustering methods will enable the cross-pollination of ideas across various research fields. We here conduct a comprehensive review of original clustering methods for functional data. We propose a systematic taxonomy that explores the connections and differences among the existing functional data clustering methods and relates them to the conventional multivariate clustering methods. The structure of the taxonomy is built on three main attributes of a functional data clustering method and therefore is more reliable than existing categorizations. The review aims to bridge the gap between the functional data analysis community and the clustering community and to generate new principles for functional data clustering.
\end{abstract}

\section{Introduction}\label{Intro}
With the advancement of data-collection technology, a wide range of industry and business sectors are now able to collect functional data. According to Ramsay and Silverman \cite{ramsay2006functional}, a functional datum is not an individual value but rather a set of measurements/observations along a continuum that, taken together, are to be regarded as a single entity. Functional data come in many forms, but their defining quality is that they consist of functions -- often, but not always, curves. For example, spectroscopic techniques obtain spectral information by probing each sample with electromagnetic radiation that varies in a range of wavelengths, and hence the calculated absorption coefficient is a function of wavelength. By probing a sample at different wavelengths, the set of absorption coefficients is one data unit. Another example of functional data is an fMRI time series, consisting of a time series of 3D images of the living human brain, where each 3D image consists of a large number of voxels (3D pixels). For example, the prevalent BOLD fMRI detects the blood-oxygen-level-dependent signal that reflects changes in deoxyhemoglobin, driven by localized changes in brain blood flow and blood oxygenation. Each 3D image is a functional datum (or, equivalently, a random field). Paradigmatic formats of functional data include time series, trajectories, spatio-temporal data, etc. However, the term ``functional'' is not the defining quality of time series, trajectories, or spatio-temporal data. Ansari et al. \cite{Ansari20202381} classified spatio-temporal data into five types, according to which certain types of spatio-temporal data are not functional data. Apart from the difference in the definitions of data format, the main difference is in the focus of statistical analysis: the focus of functional data analysis is on analyzing relations among the random elements, rather than properties of individual random elements.

While functional data analysis has received attention from statisticians since the 1980s, there is very little advancement in the area of functional data clustering. Within the two databases: Scopus and Web of Science, we found only about 100 articles that are on developing clustering methods for functional data.\footnote{In the appendix, we give the details on the identification of relevant literature and the article selection process. We also provide a table that implements the classification of the reviewed articles according to our taxonomy.} Moreover, nearly all documented methods tackle only the functional-data part of the problem, not the clustering part of the problem. For example, many studies mainly concern extracting a tabular-data proxy for functional data, ignoring the synergy between the feature-learning (a.k.a., representation-learning) step and the clustering step. The main objective of our review is to develop an overarching structure of existing functional data clustering methods, which highlights the similarities and differences among them and their connections with conventional multivariate clustering methods. We point to a few good references that give excellent coverage of state-of-the-art clustering methods for relevant data types (i.e., time series, trajectory data, and spatio-temporal data). We also suggest a new methodological framework that extricates the primary deficiency in the current tandem approach. The review will also help connect the machine learning and computer science communities with the challenges and opportunities in analyzing functional data.

\begin{figure}[!ht]
	\centering
	\includegraphics[width=12cm]{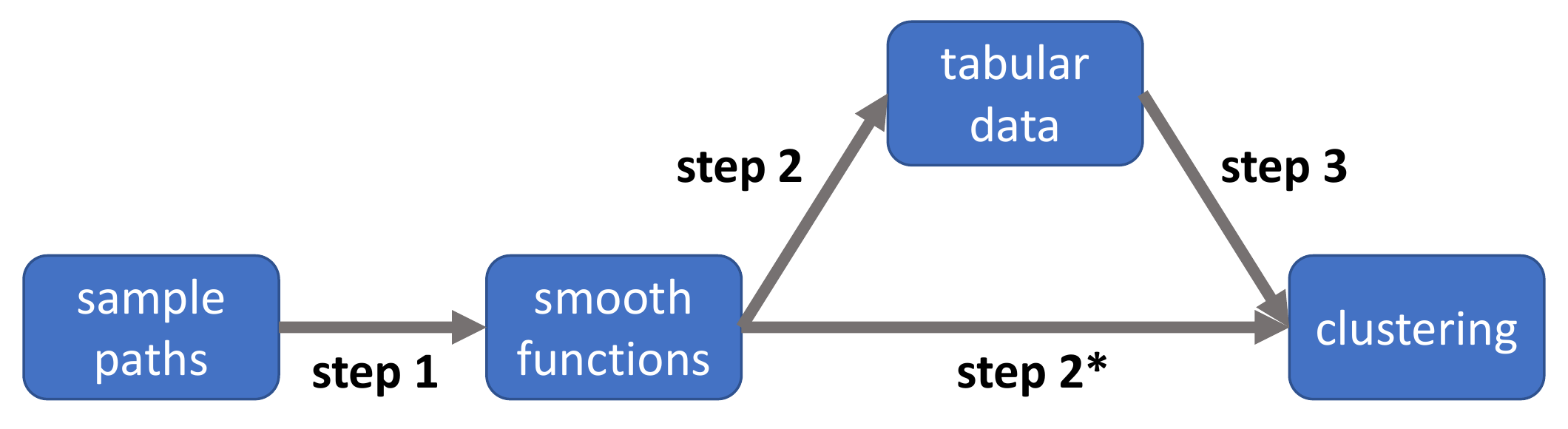}
	\caption{Functional data clustering methods can be categorized into two major groups, according to whether the clustering method is applied to the extracted tabular data (steps 1, 2 and 3) or to the estimated smooth functions (step 1 and step 2*). In the upper line approach, cluster analysis is performed in a finite-dimensional space, while in the bottom line approach, cluster analysis is performed in an infinite-dimensional space.}
\label{twostep}
\end{figure}

\begin{figure}[!ht]
	\centering
	\includegraphics[width=14cm]{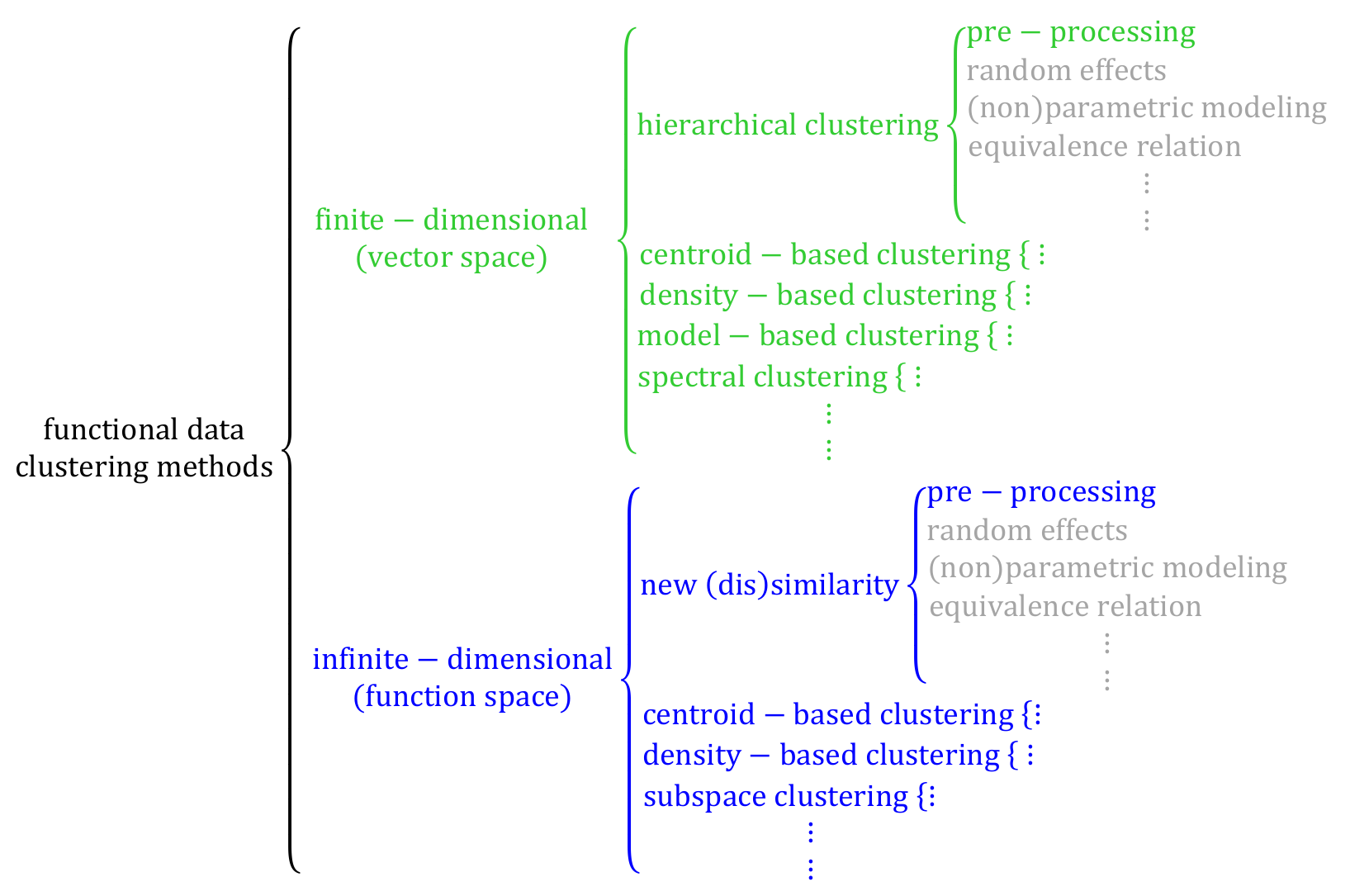}
	\caption{The three-tier categorization of existing functional data clustering methods. The first tier categorization concerns the dimension of the direct input to a clustering method, the second tier categorization is based on the characteristics of the clustering method, and the third tier categorization is to highlight the different strategies that deal with phase variation and/or amplitude variation. Methods highlighted in green and blue constitute the vast majority of the literature and are respectively reviewed in Section  \ref{VectorSpace} and Section \ref{FunctionSpace}. Methods highlighted in grey explicitly address the phase variation and/or amplitude variation in their clustering methods and are reviewed in Section  \ref{CurveRegistration}.}
\label{taxonomy}
\end{figure}

Figure \ref{twostep} 
depicts the tandem approach adopted in the current practice of functional data clustering, and Figure \ref{taxonomy} illustrates our taxonomy.
Functional data clustering methods can be categorized (Tier 1 categorization) according to whether the clustering method is applied to the extracted tabular data (i.e., in a finite-dimensional space) or to the estimated smooth functions (i.e., in an infinite-dimensional space). Then within each major category, clustering methods can be further categorized (Tier 2 categorization) according to the definition of (dis)similarity, the definition of cluster, and/or algorithmic features. In particular, in the upper pipeline of Figure \ref{twostep}, clustering methods can be classified into ``hierarchical clustering'', ``model-based clustering'', ``centroid-based clustering'', ``density-based clustering'', ``spectral clustering'', etc. In the bottom pipeline, clustering methods can be classified into ``subspace clustering'', ``nonparametric Bayesian'', ``density-based clustering'', ``new (dis)similarity'', etc. Finally, in the Tier 3 categorization, clustering methods are grouped according to the way they deal with phase variation and/or amplitude variation. In the random-effects category, phase variation and amplitude variation are characterized by a few random parameters in the function expression; for example, $y=y(at+b)$, where $t$ is the argument, and the random parameters $a$ and $b$ are to capture the phase variation. In the (non)parametric category, the time-warping functions admit either a parametric model or a nonparametric model. In the equivalence-relation category, two functions are equivalent if they can be transformed to each other by, e.g., a linear time-warping function. Our three-tier categorization provides a well-conceived and useful taxonomy in that it frames the three defining features of functional data clustering methods: dimensionality reduction, clustering strategy, and curve registration.

There are a few attempts at devising taxonomic categories for functional data clustering methods. The short survey given by Jacques and Preda \cite{Jacques2014231} classifies a few conventional functional data clustering methods into three categories. Chamroukhi and Nguyen \cite{Chamroukhi2019} reviewed a few articles that differ in the way of extracting tabular data but all apply the model-based clustering technique on the extracted tabular data. Cheam and Fredette \cite{CHEAM2020360} reviewed a few functional data clustering methods and categorized them according to whether they allow amplitude variation and/or phase variation within clusters. We note that, while a few functional data clustering methods explicitly deal with phase variation, the majority of functional data clustering methods adopt the convention that phase variation, whether relevant or not to the clustering problem, will be identified in the pre-processing step. Hence, the categories provided by \cite{CHEAM2020360} are too broad to enlighten future works. By contrast, our three-tier categorization provides a lot more information. Moreover, none of the above surveys tends to be as comprehensive as we are in this review. Ullah and Finch \cite{Ullah2013} conducted a systematic overview of applications of functional data analysis, covering all articles published during 1995 -- 2010. Cuevas \cite{CUEVAS20141} provided a good survey of the current theory and statistics of functional data analysis. Finally, while there is limited literature in the field of functional data clustering, there is abundant literature on clustering time series, trajectory data, or spatio-temporal data. Readers are referred to the following recent surveys for cross-pollination of insights and ideas: Zheng \cite{10.1145/2743025} for trajectory data, Aghabozorgi et al. \cite{AGHABOZORGI201516} for time series, and Atluri at al. \cite{10.1145/3161602}, Ansari et al. \cite{Ansari20202381} and Wang et al. \cite{9204396} for spatio-temporal data.

The novelty of functional data clustering obliges us to start by clarifying the terminology in Section \ref{Preliminaries}. The majority of the different functional data clustering methods are explained in Section \ref{VectorSpace} \& \ref{FunctionSpace}, while Section \ref{Vector-Valued} \& \ref{Dependent} are respectively dedicated to the clustering methods for vector-valued functional data and dependent functional data, which are two demanding tasks in this field. All the methods reviewed in Section \ref{VectorSpace}-\ref{Dependent} belong to the ``pre-processing'' category in Tier 3 categorization. Only a few articles, reviewed in Section \ref{CurveRegistration}, explicitly address the phase variation problem in their clustering methods. We conclude our review by presenting in Section \ref{NewFramework} a new methodological framework that aims at maximizing the synergy among the sequential steps in a functional data clustering method. The layout of our overview in each section is consistent with the hierarchy of our taxonomy. However, we may explain an original work and its follow-up or relevant works together, to avoid repeating the problem context and to provide an integrated view. Table \ref{taxonomygroup} in the appendix delineates the classification of all the reviewed publications according to our taxonomy.

\section{Preliminaries}\label{Preliminaries}
The notion ``random function'' is a natural generalization of the notion ``random variable''.  Let $\mathscr{T}$ denote a compact set in a topological space of dimension $d$ $(\geq1)$. For example, $\mathscr{T}$ can be an interval or a manifold. A \textit{random function} $Y$ is defined on a probability space $(\Omega, \mathscr{F}, \mathbb{P})$ and takes values in an infinite-dimensional space $\mathscr{Y}$. Most theoretical developments require that the sample space $\mathscr{Y}$ is $L^2 (\mathscr{T}, \mathbb{R}^p)$ -- the separable Hilbert space of all square-integrable measurable functions that are defined on $\mathscr{T}$ and taking values in $\mathbb{R}^p$ ($p\geq1$). In other words, a random function is a family of random variables $Y=\{Y(t): t\in\mathscr{T}\}$, each defined on $(\Omega, \mathscr{F}, \mathbb{P})$ and taking values in $\mathbb{R}^p$. The function $y=Y(\cdot, \omega)\in \mathscr{Y}$ is called the \textit{sample function} of the random function $Y$ at the outcome $\omega\in\Omega$. When $p\geq2$, then $Y$ is a vector-valued random function. The components of a vector-valued random function $Y$ (resp. sample function $y$) are denoted by $\{Y^1, \ldots, Y^p\}$ (resp. $\{y^1, \ldots, y^p\}$). Let $\mu(t)=\mbox{E}[Y(t)]$ denote the mean function, and $\Sigma(s, t)$ either the covariance function ($p=1$) or the matrix of variance-covariance functions ($p>1$): $\Sigma(s, t)=\mbox{E}[(Y(s)-\mu(s))(Y(t)-\mu(t))^T]$, for any $s, t\in\mathscr{T}$. Here, the superscript $T$ is the transpose operator. We might let $\Sigma$ (without any argument) denote the covariance operator of the random function $Y$. A set of \textit{functional data} is a collection of sample functions $\{y_1, \cdots, y_n\}$, each being a random realization of a random function. Note that, in this article, we treat random fields as an example of random functions.

Let the $i$th sample function $y_i$ be a realization of the random function $Y$; we often write $y_i=\mu+x_i$, and hence the sample functions $\{x_1, \cdots, x_n\}$ will have zero mean. In real practice, the observation of $y_i$ at any point $t\in\mathscr{T}$, denoted by $\tilde{y}_i(t)$, may come with an additive error: $\tilde{y}_i(t)=y_i (t)+\epsilon_i (t)$, where $\epsilon_i$ is the noise process with $\mbox{E}[\epsilon_i (t)]=0$ and $\mbox{E}[\epsilon_i^2 (t)]=\sigma^2_i(t)$. Let $\underline{t}_i=\{t_{i1}, \cdots, t_{ir_i}\}$ denote the sampling scheme for the $i$th sample function, where $t_{ij}\in\mathscr{T}$ for $j=1, \cdots, r_i$. Then the sequence of observations $\{\tilde{y}_i (t_{i1}), \cdots, \tilde{y}_i (t_{ir_i})\}$ is called a \textit{sample path} of the $i$th sample function $y_i$. Let $\mathscr{D}=\{\tilde{y}_i(\underline{t}_i): i=1, \cdots, n\}$ be an observed sample of the $n$ sample functions; that is, $\mathscr{D}$ is a set of $n$ sample paths. Here, by writing $\tilde{y}_i(\underline{t}_i)$, we employ compact notation for functions applied to collections of input points.

Clustering aims at partitioning a set of subjects into homogeneous groups, so that the subjects within a group are similar to each other and are dissimilar to any member of any other group. To mathematically define the functional data clustering problem, we assume that there are $K (\geq2)$ clusters in the population. If a sample function $y_i$ belongs to the $k$th cluster $(1\leq k\leq K)$, then it is a random realization of the $k$th random function $Y_k$ that is defined on the probability space $(\Omega, \mathscr{F}, \mathbb{P}_k)$ and taking values in $\mathscr{Y}$. In other words, there are $K$ different probability measures defined on the $\sigma$-algebra $(\Omega, \mathscr{F})$. A clustering method is to identify the underlying random function for each sample path $\tilde{y}_i(\underline{t}_i)$. Alternatively, in the popular hidden-variable formulation, the complete data are in the form of $\{(y_i, z_i): i=1, \cdots, n\}$ that are independent realizations of the couple $(Y, Z)$, where $Z$ is the hidden cluster-indicator variable with $\Pr(Z=k)=\pi_k$, $\pi_k>0$ and $\sum_{k=1}^K\pi_k=1$. Given a value of $Z$, e.g., $Z=k$, the conditional distribution of the random function $Y$ is that of $Y_k$. Note that $Y^j$ with a superscript means the $j$th component random function of $Y$, while $Y_k$ with a subscript means the $k$th random function defined on the probability space $(\Omega, \mathscr{F}, \mathbb{P}_k)$.

For the convenience of explanation, we will tacitly assume in the sequel that $d=1$ and $p=1$, unless otherwise noted. Each $\epsilon_i$ is a Gaussian white noise process, with either an individual variance $\sigma_i^2$ or a common variance $\sigma^2$. Table \ref{notations} summarizes the notation we will use throughout the work. All vectors are column vectors. Note that the bold font only applies to real-valued vectors and matrices, not vector-valued functions.
\begin{table}[!t]
  \centering
  \caption{Notation Adopted throughout the Paper.}\label{notations}
  \begin{tabular}{ll}
    \hline
    $Y$ & random function vector: $Y=(Y^1, \ldots, Y^p)^T$\\
    $\mathscr{T}$ & function domain \\
    $\{y_i: i=1, \ldots, n\}$ & $n$ sample functions\\
    $x_i$ & centered sample function $x_i=y_i-\mbox{E}[Y]$\\
    $\epsilon_i$ & noise process\\
    $\tilde{y}_i(t)$ & observation of $y_i$ at $t\in\mathscr{T}$: $\tilde{y}_i(t)=y_i(t)+\epsilon_i(t)$\\
    $\underline{t}_i=\{t_{i1}, \ldots, t_{ir_i}\}$ & sampling scheme of the $i$th sample function\\
    $\tilde{y}_i(\underline{t}_i)$ & $i$th sample path\\
    $K$ & number of clusters in the population\\
    $Y_k$ &  $k$th random function with probability space $(\Omega, \mathscr{F}, \mathbb{P}_k)$\\
    $\mu_k(t)$, $\Sigma_k(s, t)$&  mean function and covariance function of $Y_k$\\
    $\Sigma_k$ & covariance operator of $Y_k$\\
    $\pi_k$ & mixing proportion $\pi_k=\Pr(Z=k)$\\
    $z_i$ & hidden cluster label of $y_i$: $z_i\in\{1, \ldots, K\}$\\
    $\langle\cdot, \cdot\rangle$, $\|\cdot\|$ & inner product and norm of $L^2(\mathscr{T}, \mathbb{R}^p)$\\
    $\|\cdot\|_2$ & Euclidean norm for vectors\\
    $\circ$ & function composition operator\\
    $\delta(\cdot)$ & Dirac delta function\\
    $D^vy$ & derivative of order $v$ of a function $y$\\
    $\{b_1, b_2, \ldots\}$ & generic notation for a set of basis functions\\
    diag($\cdot$) & diagonal matrix of the argument\\
    $\mathscr{N}(\cdot, \cdot)$ & multivariate Gaussian distribution\\
    \hline
  \end{tabular}
\end{table}

\section{Clustering in Finite-Dimensional Space}\label{VectorSpace}
B-splines, wavelets and functional principal component (fPC) decomposition are the three prevalent smoothing techniques. We here briefly explain the fPC decomposition and wavelet smoothing. With the Karhunen-Lo\`{e}ve theorem, we can decompose the covariance function $\Sigma(s, t)$ of $Y$ into
\begin{equation*}
\Sigma(s, t)=\sum_{v=1}^{\infty}\lambda_v b_v(s)b_v(t),
\end{equation*}
where $\lambda_1\geq\lambda_2\geq\cdots\geq0$ are the eigen-values, and $\{b_v\}_{v\in\mathbb{N}}$ are the orthonormal eigen-functions. Then any sample function $y_i$, a realization of $Y$, has the following expansion
\begin{equation*}
y_i(t)=\mu(t)+\sum_{v=1}^{\infty}a_{iv} b_v(t),
\end{equation*}
where $a_{iv}=\langle y_i-\mu, b_v\rangle$ is the fPC score associated with the eigen-function $b_v$, with the properties E$[a_{iv}]=0$, var$(a_{iv})=\lambda_v$, and E$[a_{iv}a_{ir}]=0$ for $v\neq r$. Note that the fPC scores $\{a_{i1}, a_{i2}, \ldots\}$ are always uncorrelated, but the independence is only guaranteed when $Y=\{Y(t): t\in\mathscr{T}\}$ is a Gaussian process.

The discrete wavelet transform framework utilizes two sets of orthonormal basis functions $\{\beta_{jv}(t)=\sqrt{2^j}\beta(2^jt-v): j\in\mathbb{N}, v\in\mathbb{Z}\}$ and $\{b_{jv}(t)=\sqrt{2^j}b(2^jt-v): j\in\mathbb{N}, v\in\mathbb{Z}\}$, i.e., scaled and translated replicas of the scaling function $\beta$ and the wavelet function $b$. Wavelet expansion is defined in terms of a sequence of nested closed subspaces $V_j$ of the space $L^2(\mathscr{T}, \mathbb{R})$: $L^2(\mathscr{T}, \mathbb{R})\supset\cdots\supset V_1\supset V_0\supset V_{-1}\supset \cdots$. For each resolution level $j$, the space $V_j$ is spanned by $\{\beta_{j v}\}_{v\in\mathbb{Z}}$; the orthogonal complement $W_j$ of $V_j$ with respect to $V_{j+1}$  is spanned by $\{b_{j v}\}_{v\in\mathbb{Z}}$. Repeating the direct-sum decomposition, we have $V_J=V_{J-1}\oplus W_{J-1}=V_j\oplus W_j\oplus \cdots\oplus W_{J-1}$, for any lowest resolution $j$ of interest ($0\leq j\leq J-1$). If we assume $y\in V_J$, it follows from the decomposition that
\begin{equation*}
  y(t)=\sum_{v\in\mathbb{Z}}\alpha_{j v} \beta_{j v}(t)+\sum_{r=j}^{J-1}\sum_{v\in\mathbb{Z}}a_{rv} b_{rv}(t),
\end{equation*}
where $\alpha_{j v}$ and $a_{rv}$ are respectively the scale and wavelet coefficients. When the function $y$ is evaluated on a uniform lattice of a compact interval, say, a grid of $2^J$ lattice points on the interval $[0, 1]$, then it is natural to assume that $y\in V_J$ and write $y(t)=\alpha_{00}\beta_{00}(t)+\sum_{r=0}^{J-1}\sum_{v=0}^{2^r-1}a_{rv} b_{rv}(t)$.

\begin{figure}[!ht]
	\centering
	\includegraphics[width=14cm]{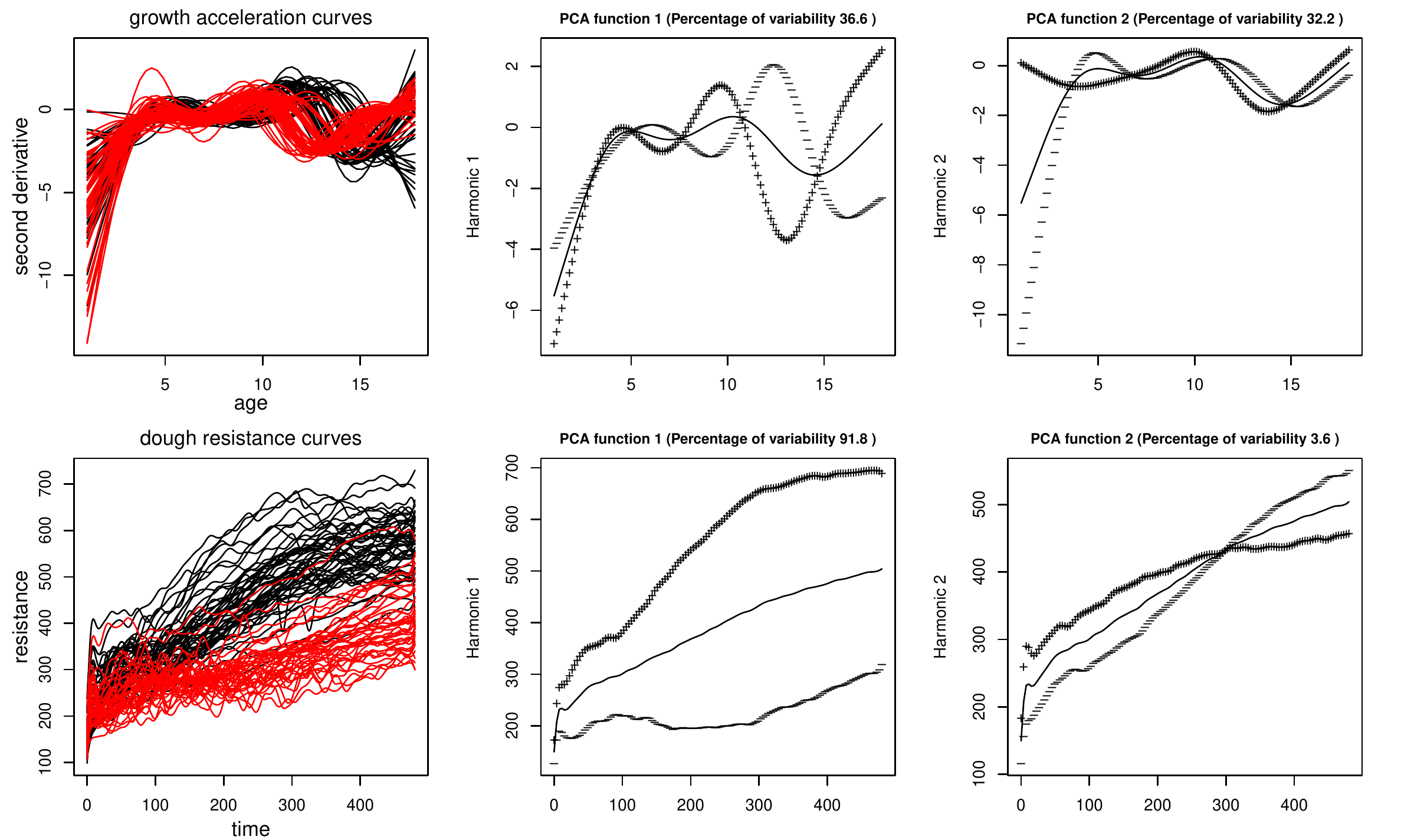}
	\caption{Top left: Growth acceleration curves of 54 girls (in red) and 39 boys (in black) from the Berkeley growth study. Bottom left: Dough resistance curves of 40 bad-quality flours (in red) and 50 good-quality flours (in black) during the kneading process. The middle (resp. right) panels show the mean function and the effects of adding ($+$) and subtracting ($-$) a suitable multiple of the first (resp. second) principal component function. For the growth acceleration curves, the first four principal component functions respectively explain 36.6\%, 32.2\%, 24.3\% and 2.8\% variation in the functions. For the dough resistance curves, the first principal component function alone explains 91.8\% variation in the functions.}
\label{example_pca}
\end{figure}

Figure \ref{example_pca} 
gives two examples of fPC decomposition. For the growth acceleration curves, the original data are the heights of 54 girls and 39 boys measured at 31 stages from 1 to 18 years of age. The data are from the Berkeley growth study and available in the R package \textit{fda}. In line with the study in \cite{ramsay2006functional}, we here focus on the growth acceleration curves, namely the second derivatives of the estimated smooth functions. For the dough resistance curves, the original data are the dough resistance records of 40 bad-quality flours, 25 medium-quality flours, and 50 good-quality flours during the kneading process, measured every 2 seconds over a time period of 480 seconds. The data are from Danone Vitapole Paris Research Center, and the data used for Figure \ref{example_pca} are available at \url{http://math.univ-lille1.fr/~preda/FDA/flours.txt}. The fPC decomposition was performed on the estimated smooth functions. For all datasets except the Canadian weather dataset, the B-spline smoothing technique is adopted for obtaining the estimated smooth functions, where the size of the smoothing parameter is determined by the generalized cross validation (see Chapter 5 of \cite{ramsay2006functional}). For the Canadian weather dataset, we applied the Fourier basis for smoothing.

\subsection{Model-Based Clustering}
A branch of works adopts the mixed-effects model: given $z_i=k$, write $y_i(t)=\mu_k(t)+x_i(t)=\sum_{v=1}^m\alpha_{kv}b_v(t)+\sum_{v=1}^ma_{iv}b_v(t)$, where $\{b_1, \ldots, b_m\}$ are pre-determined basis functions; the mean function $\mu_k$ is treated as the fixed effect, while $x_i$ is treated as the random effect. The pioneering work \cite{JamesSugar} assumes that each random function $Y_k=\{Y_k(t):t\in\mathscr{T}\}$ is a Gaussian process, having an individual mean function $\mu_k$ yet a common covariance function. Therefore, conditioned on the hidden cluster labels $\{z_1, \cdots, z_n\}$, the centered functions $\{x_1, \cdots, x_n\}$ are i.i.d. realizations from a zero-mean Gaussian process. Taking the mixed-effects modeling approach with cubic B-spline basis functions, they assumed that the i.i.d. coefficient vectors $\pmb{a}_i=(a_{i1}, \ldots, a_{im})^T$ have a zero-mean Gaussian distribution. An EM-type algorithm was developed for parameter estimation. Giacofci et al. \cite{Giacofci201331} replaced the cubic B-spline basis with a wavelet basis, and Nguyen et al. \cite{Nguyen201676} replaced the cubic B-spline basis with the linear nodal basis to deal with the case when $d=2$; both works assume that, conditioned on $z_i=k$, the random-effect coefficient vector $\pmb{a}_i$ has a zero-mean Gaussian distribution with a cluster-specific diagonal covariance matrix. Nguyen et al. \cite{Nguyen20185} applied the cubic B-spline basis expansion directly on $y_i$: $y_i(t)=\sum_{v=1}^ma_{iv}b_v(t)$, and estimated each coefficient vector $\pmb{a}_i$ separately via the ordinary least squares regression technique. They then assumed that the estimated coefficient vectors have the traditional Gaussian mixture model. A variant given by Ma and Zhong \cite{Ma2008625} adopts the functional ANOVA decomposition of $\mu_k$: with $d\geq2$, $\pmb{t}=(t^1, \ldots, t^d)$, and $y_i=\mu_k+x_i$, the formulation of $x_i$ is $x_i(\pmb{t})=\sum_{v=1}^ma_{iv}b_v(\pmb{t})$, and of $\mu_k$ is $\mu_k(\pmb{t})=u_{k0}+\sum_{v=1}^d\mu_{kv}(t^v)+\sum_{v=1}^d\sum_{r=v+1}^d\mu_{kvr}(t^v, t^r)+\cdots$. They assumed that the $\mu_k$'s are from a reproducing kernel Hilbert space, and that the random-effect coefficient vector $\pmb{a}_i$ has a zero-mean Gaussian distribution with a cluster-specific covariance matrix. The penalized Henderson's likelihood was employed for parameter estimation, and utilizing the representer theorem, each mean function $\mu_k$ can be expressed as a linear combination of a few basis functions.

Another branch of works adopts the notion of subspace. Let $\mathscr{B}=\{b_1 , \ldots, b_m\}$ denote a set of $m$ pre-determined basis functions. Then each cluster is a linear space spanned by a subset of basis functions from $\mathscr{B}$. Given $z_i=k$, let $\pmb{a}_i$ denote the projection coefficient vector of $y_i$ onto the space spanned by $\{b_1 , \ldots, b_m\}$, and $\pmb{\alpha}_i$ the projection coefficient vector of $y_i$ onto the $k$th subspace. Bouveyron and Jacques \cite{Bouveyron2011281} assumed that, for the $k$th cluster/subspace, there exists an $m\times m$ orthogonal matrix $\pmb{Q}_k=[\pmb{W}_k, \pmb{V}_k]$, with $\pmb{W}_k=[w^k_{vr}]_{m\times m_k}$, such that the basis functions for the $k$th subspace are $\{\sum_{v=1}^mw^k_{v1}b_v, \ldots, \sum_{v=1}^mw^k_{vm_k}b_v\}$. They further made the following distributional assumptions: (1) $\pmb{a}_i=\pmb{W}_k\pmb{\alpha}_i+\pmb{e}_i$, (2) the distribution of $\pmb{\alpha}_i$ is $\mathscr{N}(\pmb{u}_k, \mbox{diag}(\pmb{\gamma}_k))$, and of $\pmb{e}_i$ is $\mathscr{N}(\pmb{0}, \pmb{\Upsilon}_k)$, (3) the distribution of $\pmb{a}_i$ is $\mathscr{N}(\pmb{W}_k\pmb{u}_k, \pmb{\Gamma}_k)$, where $\pmb{\Gamma}_k=\pmb{W}_k\mbox{diag}(\pmb{\gamma}_k)\pmb{W}_k^T+\pmb{\Upsilon}_k$, and (4) the covariance matrix $\pmb{\Upsilon}_k$ satisfies the constraint that $\pmb{Q}_k^T\pmb{\Gamma}_k\pmb{Q}_k=\mbox{diag}(\pmb{\gamma}_k, \gamma_{k0}, \ldots, \gamma_{k0})$. An EM-type algorithm was developed to estimate all unknown parameters. The work \cite{Sharp2021735} differs from \cite{Bouveyron2011281} mainly in the distributional assumptions: the distribution of $\pmb{\alpha}_i$ is a generalized hyperbolic distribution, and the distribution of $\pmb{a}_i$ is another generalized hyperbolic distribution. The work \cite{10.1214/15-AOAS861} aims at finding one subspace, generated by the orthogonal matrix $\pmb{Q}=[\pmb{W}, \pmb{V}]$, such that the clusters of the projected coefficient vectors are well separated. They made the following distributional assumptions: given $z_i=k$, (1) $\pmb{a}_i=\pmb{W}\pmb{\alpha}_i+\pmb{e}_i$, (2) the distribution of $\pmb{\alpha}_i$ is $\mathscr{N}(\pmb{u}_k, \pmb{\Upsilon}_k)$, and of $\pmb{e}_i$ is $\mathscr{N}(\pmb{0}, \pmb{\Upsilon}_0)$, (3) the distribution of $\pmb{a}_i$ is $\mathscr{N}(\pmb{W}\pmb{u}_k, \pmb{\Gamma}_k)$, where $\pmb{\Gamma}_k=\pmb{W}\pmb{\Upsilon}_k\pmb{W}^T+\pmb{\Upsilon}_0$, and (4) the covariance matrix $\pmb{\Upsilon}_0$ satisfies the constraint that $\pmb{Q}^T\pmb{\Gamma}_k\pmb{Q}=\mbox{diag}(\pmb{\Upsilon}_k, \gamma_0, \ldots, \gamma_0)$.

Utilizing the notion of density defined by Delaigle and Hall \cite{Delaigle20101171}, both Jacques and Preda \cite{JACQUES2013164} and Rivera-Garc\'{\i}a et al. \cite{Rivera-Garca2019201} assumed that the random functions are Gaussian, and therefore the random coefficients in the fPC decomposition are independent and Gaussian distributed. Given $\Sigma_k(s, t)=\sum_{v=1}^{\infty}\lambda_{kv} b_{kv}(s)b_{kv}(t)$, $z_i=k$, and $y_i(t)=\mu_k(t)+\sum_{v=1}^{m_k}a_{ikv} b_{kv}(t)$, Jacques and Preda \cite{JACQUES2013164} assumed that the distribution of the coefficient vector $(a_{ik1}, \ldots, a_{ikm_k})^T$ is $\mathscr{N}(\pmb{0}, \mbox{diag}(\lambda_{k1}, \ldots, \lambda_{km_k}))$. An EM-type algorithm was developed, in which the M step involves updating the fPC decomposition and determining the number of principal components for each cluster. Rivera-Garc\'{\i}a et al. \cite{Rivera-Garca2019201} assumed that the distribution of the coefficient vector $(a_{ik1}, \ldots, a_{ikm})^T$, with $m>m_k$, is $\mathscr{N}(\pmb{0}, \mbox{diag}(\lambda_{k1}, \ldots, \lambda_{km_k}, \lambda_k, \ldots, \lambda_k))$; that is, the additional ($m-m_k$) random coefficients are identically distributed. The work \cite{Bouveyron20151143} differs from \cite{Rivera-Garca2019201} mainly in that Bouveyron et al. \cite{Bouveyron20151143} assumed that the transformed random function $g(Y_k)$ is a Gaussian process, with mean function $\mu_k$ and covariance function $\Sigma_k(s, t)$, and the function $g$ is the feature map of a kernel function.

The basis functions in Wu et al. \cite{WuWang2021} are the eigen-functions for the covariance operator of $\sum_{k=1}^K\pi_kY_k$; that is, the Karhunen-Lo\`{e}ve expansion is performed on the covariance function estimated from all the functional data. The extracted fPC scores are the tabular-data proxy. They assumed that the distribution of the cluster members around their projections onto the cluster's principal curve is an isotropic Gaussian, and that the projections of cluster members onto the cluster's principal curve are uniformly distributed. A principal curve for a dataset is a one-dimensional curve that passes through the middle of the given data: if we pick any point on the curve, collect all of the data that project onto this point, and average them, then this average coincides with the point on the curve. Therefore, given $z_i=k$, the density function of the proxy $\pmb{a}_i$ is $\frac{1}{l_k}\times\frac{1}{\sqrt{2\pi}\gamma_k}\exp(-\frac{\|\pmb{a}_i-\mathscr{P}_k(\pmb{a}_i)\|_2^2}{2\gamma_k^2})$, where $l_k$ is the length of the $k$th principal curve, and $\mathscr{P}_k(\pmb{a}_i)$ is the projection of $\pmb{a}_i$ onto the $k$th principal curve. An EM-type algorithm was developed, where the M step includes updating the principal curve for each cluster.

\subsection{Centroid-Based Clustering}
With the popularity of centroid-based clustering methods, including the $k$-means and $k$-medoids techniques, a generic methodological framework is to apply a centroid-based clustering method on the extracted basis-expansion coefficient vectors. We should mention that, as noted by Tarpey and Kinateder \cite{Tarpey2003} and Tarpey \cite{Tarpey200734}, if the basis is not orthogonal, then applying the multivariate $k$-means algorithm on the coefficient vectors is not equivalent to the functional $k$-means algorithm with the $L^2$ distance metric.

Abraham et al. \cite{Abraham2003} coupled a B-spline basis and the $k$-means clustering algorithm. Garc\'{\i}a-Escudero and Gordaliza \cite{Garcia-Escudero2005185} adopted the cubic B-spline basis and a variant of the $k$-means algorithm, called trimmed $k$-means algorithm, where in each iteration of the algorithm, a fixed number of the most outlying coefficient vectors are excluded from calculating the cluster centers. Giordani et al. \cite{journal.pone.0242197} coupled a B-spline basis and the fuzzy $k$-medoids clustering algorithm. Denis et al. \cite{10.1111/rssc.12404} assumed that the sample functions are piecewise linear and applied a second-order B-spline basis for smoothing. The proxy for each sample function is the vector of knots and coefficients from the smoothing spline, and the $k$-means algorithm was applied on the extracted tabular data. Kim and Oh \cite{KIM2020104626} adopted a B-spline basis and the $k$-means algorithm. However, the smoothing technique was applied, not on the mean function, but on the quantile functions: $y_i(t; \tau)=\sum_{v=1}^ma_{iv}^\tau b_v(t)$, where $y_i(t; \tau)$ is the $\tau$th quantile curve of $y_i$, estimated from the sample path $\tilde{y}_i(\underline{t}_i)$. The proxy for the sample function $y_i$ is the weighted average of $\{\pmb{a}_i^{\tau_1}, \pmb{a}_i^{\tau_2}, \ldots\}$, the quantile-smoothing coefficient vectors at a few different quantile levels.

Both Antoniadis et al. \cite{Antoniadis2013} and Lim et al. \cite{Lim2019368} utilized the multi-resolution aspect of the discrete wavelet transform: $y(t)=\alpha_{00}\beta_{00}(t)+\sum_{r=0}^{J-1}\sum_{v=0}^{2^r-1}a_{rv} b_{rv}(t)$. Because all basis functions are orthonormal, we have the energy decomposition $\|y\|^2=\alpha_{00}^2+\sum_{r=0}^{J-1}\|\pmb{a}_r\|_2^2$. In Antoniadis et al. \cite{Antoniadis2013}, the energies $\{\alpha_{00}^2, \|\pmb{a}_0\|_2^2, \ldots, \|\pmb{a}_{J-1}\|_2^2\}$ are the extracted features, and the related tabular data are fed into the $k$-means algorithm. With the wavelet expansion of the difference: $y_i(t)-\mu_k(t)=\sum_{v\in\mathbb{Z}}\langle y_i-\mu_k, \beta_{0 v}\rangle \beta_{0 v}(t)+\sum_{r=0}^{J-1}\sum_{v\in\mathbb{Z}}\langle y_i-\mu_k, b_{rv}\rangle b_{rv}(t)$, Lim et al. \cite{Lim2019368} approximated the squared distance $\|y_i-\mu_k\|^2$ by the total energy $\sum_{v\in\mathbb{Z}}\langle y_i-\mu_k, \beta_{0 v}\rangle^2+\sum_{r=0}^{J-1}\sum_{v\in\mathbb{Z}}\langle y_i-\mu_k, b_{rv}\rangle^2$. For each resolution level $j$ from 1 to $J$, given the approximating distances $\{\sum_{v\in\mathbb{Z}}\langle y_i-\mu_k, \beta_{0 v}\rangle^2+\sum_{r=0}^{j}\sum_{v\in\mathbb{Z}}\langle y_i-\mu_k, b_{rv}\rangle^2\}$, they applied the functional $k$-means algorithm and divided each cluster from the resolution level $j-1$ into smaller clusters, resulting in a divisive hierarchical clustering method. With the assumption of only two clusters, Delaigle et al. \cite{Delaigle2019271} adopted the Haar basis and modified the objective function of the $k$-means algorithm into $\min_{\{z_i\}_{i=1}^n}\sum_{k=1}^{K}\sum_{z_i=k}\|\mbox{diag}(\frac{1}{n-1}\sum_{i=1}^n(\pmb{a}_i-\bar{\pmb{a}})^2)^{-1}(\pmb{a}_i-\bar{\pmb{a}}_k)\|_2^2$, where $\bar{\pmb{a}}$ is the overall sample average, and $\bar{\pmb{a}}_k$ is the cluster-wise sample average.

In Garc\'{\i}a et al. \cite{LUZLOPEZGARCIA2015231}, the sampling schemes $\{\underline{t}_i\}_{i=1}^n$ are identical with $\underline{t}=\{t_1, \ldots, t_r\}$, and the function space is a reproducing kernel Hilbert space with a Mercer kernel $\kappa(\cdot, \cdot)$. Given the two sample paths, $\tilde{y}_i(\underline{t})$ and $\tilde{y}_j(\underline{t})$, and utilizing the representer theorem, the smooth estimates of $y_i$ and $y_j$ admit the form: $\hat{y}_i(t)=\sum_{v=1}^ra_{iv}\kappa(t, t_v)$ and $\hat{y}_j(t)=\sum_{v=1}^ra_{jv}\kappa(t, t_v)$. Then the squared $L^2$ distance between $y_i$ and $y_j$ can be approximated by $(\pmb{a}_i-\pmb{a}_j)^T\kappa(\underline{t}, \underline{t})(\pmb{a}_i-\pmb{a}_j)=\|\pmb{U}\pmb{a}_i-\pmb{U}\pmb{a}_j\|_2^2$, where $\pmb{U}^T\pmb{U}=\kappa(\underline{t}, \underline{t})$. The $k$-means algorithm was repeatedly applied on data of the form $\{\pmb{W}\pmb{U}\pmb{a}_1, \ldots, \pmb{W}\pmb{U}\pmb{a}_n\}$, where the projection matrix $\pmb{W}\in\mathbb{R}^{q\times r}$, with $q<r$, was randomly generated for dimension reduction. Additionally, according to Mercer's theorem, the reproducing kernel has the representation: $\kappa(s, t)=\sum_{v=1}^{\infty}\lambda_v b_v(s)b_v(t)$, where $\lambda_1\geq\lambda_2\geq\cdots\geq0$ are the eigen-values, and $\{b_v\}_{v\in\mathbb{N}}$ are the orthonormal eigen-functions. In analogy with the fPC decomposition, the smooth estimate of $y_i$ admits the form: $\hat{y}_i(t)=\sum_{v=1}^m\alpha_{iv}b_v(t)$, where $m$ is the rank of the matrix $\kappa(\underline{t}, \underline{t})$. Then the squared distance between $y_i$ and $y_j$ can be approximated by $\sum_{v=1}^m\lambda_v^{-1}(\alpha_{iv}-\alpha_{jv})^2$. While the two approximations are equivalent when $m=r$, the dimension reduction probem is trivial in the second approach, by simply controlling the number of eigen-functions in the approximation.

%Given any set of evaluation locations $\underline{t}=\{t_1, \ldots, t_r\}$, Delaigle et al. \cite{Delaigle2012299} ran the traditional $k$-means algorithm on the tabular data $\{y_i(\underline{t}): i=1, \ldots, n\}$ to obtain a partition of the functional data: $\{\hat{z}_i: i=1, \ldots, n\}$, where $\hat{z}_i$ is the estimated cluster label of $y_i$. They then calculated the within-cluster variation of the functional data: $\sum_{k=1}^K\sum_{\hat{z}_i=k}\|y_i-\bar{y}_k\|$, where $\bar{y}_k$ is the sample average of the functions with $\hat{z}_i=k$. The objective is to determine the optimal set of evaluation locations to minimize the within-cluster variation of the functional data.

\subsection{Nonparametric Bayesian}
With the discrete wavelet transform $y_i(t)=\alpha_{i00}\beta_{00}(t)+\sum_{r=0}^{J-1}\sum_{v=0}^{2^r-1}a_{irv} b_{rv}(t)$, Ray and Mallick \cite{RayMallick} assumed that the distribution of $\{a_{irv}: i=1, \dots, n, v=0, \ldots, 2^r-1\}$ is a mixture of $\mathscr{N}(0, \gamma_r)$ and $\delta_0$ (a point mass distribution at 0), and developed the following hierarchical Bayesian framework. The coefficient vector $\pmb{a}_i=(a_{i00}, a_{i10}, a_{i11}, \ldots)^T$ and the variance $\sigma_i^2=\mbox{E}[\epsilon_i^2 (t)]$ are generated from a random distribution $G$ that is a realization from a Dirichlet process (with base distribution $G^0$). The base distribution $G^0$ is formulated by a Bayesian model: the prior on $\{\sigma_i^2\}_{i=1}^n$ is an inverse gamma distribution, the prior on $\{\pmb{a}_i\}_{i=1}^n$ is $\mathscr{N}(\pmb{0}, \mbox{diag}(\tau_{00}\gamma_0, \tau_{10}\gamma_1, \tau_{11}\gamma_1, \ldots, \tau_{rv}\gamma_r, \ldots))$, the Bernoulli random variable $\tau_{rv}$ ($v=0, \ldots, 2^r-1$) takes the value 1 with probability $\theta_r$, and finally the prior on each $\gamma_r$ is a different inverse gamma distribution. Hence, the number of clusters is not a constant but randomly determined in the P\'{o}lya urn sampling scheme, and its expected value is controlled by the concentration parameter in the Dirichlet process. Suarez and Ghosal \cite{10.1214/14-BA925} applied the nonparametric Bayesian technique to evaluate the similarity between two sample functions. They assumed that $\mbox{E}[\epsilon_i^2 (t)]=\sigma^2$ and the sequence $\{a_{1rv}, \ldots, a_{nrv}\}$ are generated from a random distribution $G_{rv}$ that is a realization from a Dirichlet process (with an individual base distribution $G^0_{rv}$ and a common concentration  parameter). The base distribution $G^0_{rv}$ is a mixture of $\mathscr{N}(0, \gamma_r)$ and $\delta_0$, and the prior on $\sigma^2$ is an inverse gamma distribution. A pairwise similarity matrix was constructed for clustering, where the similarity measure between $y_i$ and $y_j$ is $(2^J-1)^{-1}\sum_{r=0}^{J-1}\sum_{v=0}^{2^r-1}\delta(a_{irv}=a_{jrv})$.

\begin{figure}[!ht]
	\centering
	\includegraphics[width=14cm]{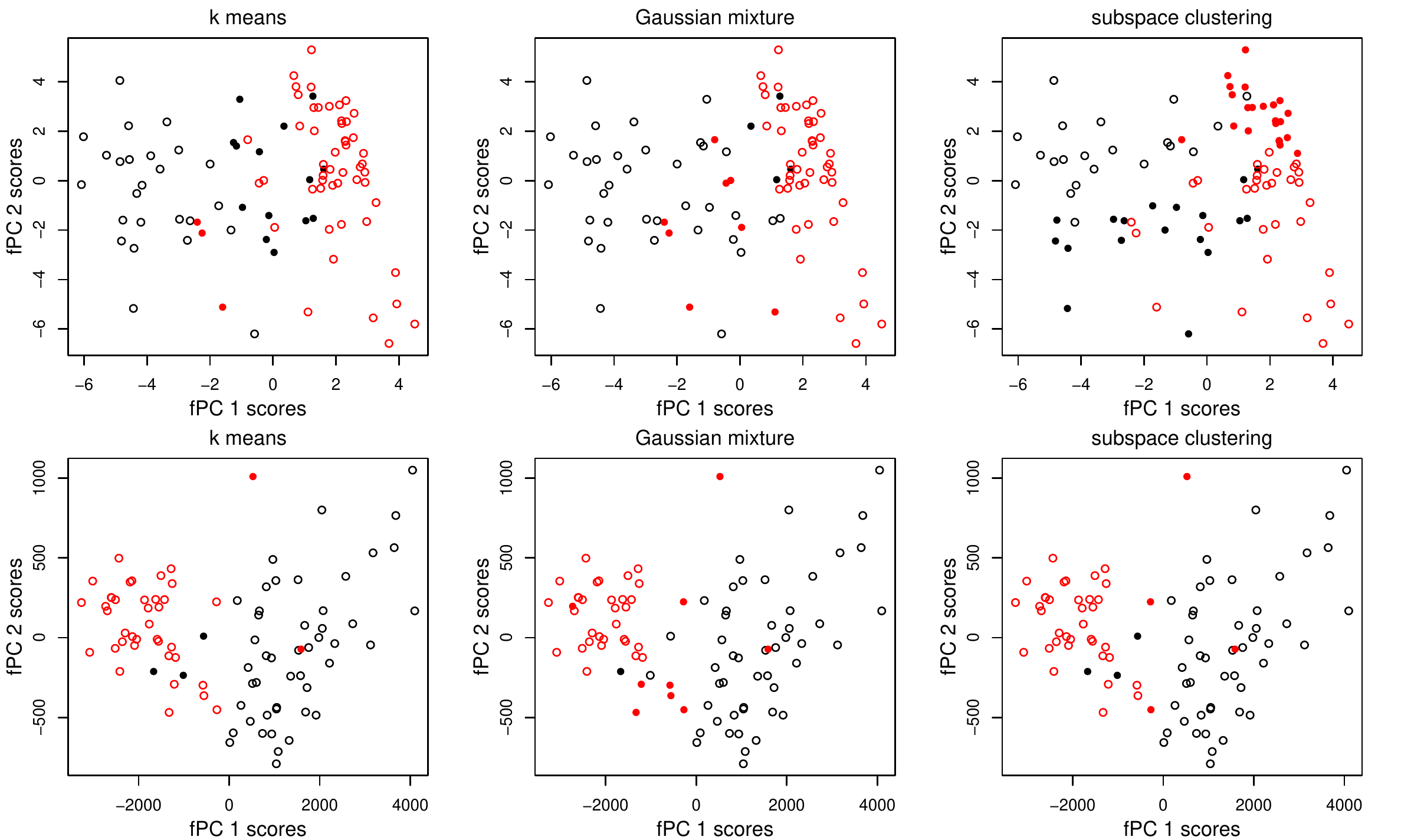}
	\caption{We apply the $k$-means algorithm (left) and Gaussian mixture model (middle) on the vectors of fPC scores, and a subspace clustering method (right) on the vectors of B-spline coefficients. The upper panels correspond to the growth acceleration curves, and the lower panels correspond to the dough resistance curves. Incorrectly clustered data points are indicated by filled circles. For example, a filled red circle in the top left panel means that a girl's curve was assigned into the boy's cluster. Here, no clustering method dominates another, and all the three methods are inadequate for the growth acceleration curves.}
\label{example_VS}
\end{figure}

The fPC decomposition in Figure \ref{example_pca} indicates that, for both datasets, the first four principal component functions together explain more than  95\% variation in the functions. Hence, an intuitive approach is to represent each curve by the vector of its first four fPC scores. Figure \ref{example_VS} 
depicts the clustering results obtained by applying the $k$-means algorithm (left) and Gaussian mixture model (middle) on the fPC score vectors. Another intuitive approach is to represent each curve by the vector of its B-spline coefficients. The dimension of the B-spline coefficient vector is much higher than that of the fPC score vector. Hence, a subspace clustering method would be more appropriate here. The clustering results in the right panels were obtained by the ``funFEM'' function in the R package \textit{funFEM}, which is built on the work \cite{10.1214/15-AOAS861}. For ease of comparison, all clustering results are visualized in the linear space of the first two principal component functions. A filled circle means that the relevant curve is incorrectly clustered.

\section{Clustering in Infinite-Dimensional Space}\label{FunctionSpace}
An intuitive approach to perform cluster analysis in a function space is to define a (dis)similarity measure for the functions in the function space.  A standard dissimilarity measure for the Hilbert space $L^2 (\mathscr{T}, \mathbb{R}^p)$ is $\mathbbm{d}_v(y, \eta)=\|D^vy-D^v\eta\|$, where $\|\cdot\|$ is the $L^2$ norm:
\begin{equation*}
\|D^vy-D^v\eta\|^2=\sum_{j=1}^{p}\int_{\mathscr{T}}[D^vy^j(t)-D^v\eta^j(t)]^2dt,
\end{equation*}
and $D^vy$ is the $v$th-order derivative of $y$. When $v\geq1$, the semi-metric $\mathbbm{d}_v(y, \eta)$ measures the dissimilarity between the derivatives of $y$ and $\eta$. Ferreira and Hitchcock \cite{FerreiraHitchcock2009} compared four agglomerative hierarchical clustering methods via simulated data, where the dissimilarity measure is $\mathbbm{d}_0$. They found that both function shape and cluster relative size affect the performance of the four clustering methods. Chen et al. \cite{biom.12161Chen} defined a weighted distance in the form of $\|y-\eta\|^2_w=\int_{\mathscr{T}}w(t)[y(t)-\eta(t)]^2dt$, where $w$ is a non-negative weighting function satisfying $\int_{\mathscr{T}}w(t)dt=1$. Assuming that the coefficient vector in the basis-expansion of any pairwise difference $y_i-y_j=\sum_{v=1}^ma_m b_v(t)$ has a multivariate Gaussian distribution, the weighting function $w$ is obtained by minimizing the coefficient of variation $\frac{\sum_{i\neq j}\mbox{var}(\|y_i-y_j\|^2_w)}{\sum_{i\neq j}\mbox{E}^2[\|y_i-y_j\|^2_w]}$. A relevant work \cite{FLORIELLO20171} introduces a weighting function to the clustering objective function: the clustering optimization problem is $\min\limits_{w, \{z_i\}_{i=1}^n}\int_{\mathscr{T}}w(t)f(y_1(t), \ldots, y_n(t); \{z_i\}_{i=1}^n)dt$, where the constraints on the weighting function are $\|w\|\leq1$, and the Lebesgue measure of the set $\{t\in\mathscr{T}: w(t)=0\}$ is larger than a threshold. When $f$ measures the within-cluster variation: $f(y_1(t), \ldots, y_n(t); \{z_i\}_{i=1}^n)=\sum_{k=1}^K\sum_{z_i=k}[y_i(t)-\bar{y}_k(t)]^2$, the framework reduces to the $k$-means technique with the weighted distance $\|y_i-\bar{y}_k\|_w^2=\int_{\mathscr{T}}w(t)[y_i(t)-\bar{y}_k(t)]^2dt$. Gaetan et al. \cite{Gaetan2016964} defined the distance between two functions via their quantile curves: the squared distance between $y_i$ and $y_j$ is $\int_0^1\int_{\mathscr{T}}[y_i(t; \tau)-y_j(t; \tau)]^2dtd\tau$, where $y_i(t; \tau)$ (resp. $y_j(t; \tau)$) is the $\tau$th quantile curve of $y_i$ (resp. $y_j$). The pairwise-distance matrix is the input to the $k$-medoids algorithm. Tzeng et al. \cite{Tzeng20183492} adopted a spline basis for smoothing and formulated the smoothed curve as a function of the roughness-penalty parameter: $\hat{y}_i(\cdot; \theta_i)=\min_{y, \theta}\{\frac{1}{r_i}\|\tilde{y}_i(\underline{t}_i)-y(\underline{t}_i)\|^2_2+\theta\int_{\mathscr{T}}[D^2y(t)]^2dt\}$. The dissimilarity between $y_i$ and $y_j$ is $\frac{1}{2}(\|\hat{y}_i(\cdot; \theta_i)-\hat{y}_j(\cdot; \theta_i)\|+\|\hat{y}_i(\cdot; \theta_j)-\hat{y}_j(\cdot; \theta_j)\|)$. The data in van Delft and Dette \cite{vanDelft2021469} are non-stationary time series, and hence the second order structure is time-dependent. Let $||\cdot||_{S_p}$ denote the the p-Schatten norm, and when $p=2$, $||\cdot||_{S_2}$ is the Hilbert-Schmidt norm. They defined the similarity measure for locally stationary time series through the (time-varying) spectral density operator: $\mathbbm{s}(y_i, y_j)=\frac{\int_{\mathscr{T}}\int_{-\pi}^\pi ||\kappa_{\upsilon, \omega}^i-\kappa_{\upsilon, \omega}^j||_{S_2}^2 d\omega d\upsilon}{\int_{\mathscr{T}}\int_{-\pi}^\pi (||\kappa_{\upsilon, \omega}^i||_{S_2}^2+||\kappa_{\upsilon, \omega}^j||_{S_2}^2) d\omega d\upsilon}$, where $\kappa_{\upsilon, \omega}^i$ (resp. $\kappa_{\upsilon, \omega}^j$) is the time-varying spectral density operator for $y_i$ (resp. $y_j$). The traditional spectral clustering technique was applied on the pairwise similarity matrix.

The Hilbert-Schmidt norm and its generalization, the p-Schatten norm, were also adopted by Ieva et al. \cite{Ieva20161} and Kashlak et al. \cite{Kashlak2019214} for the analysis of covariance operators, utilizing the property that the covariance operator of each $Y_k$ is trace-class, self-adjoint, and compact. Ieva et al. \cite{Ieva20161} assumed that the population has only two random functions, $Y_1$ and $Y_2$, that differ in their covariances, rather than in their means. The dissimilarity between the two covariance operators, $\Sigma_1$ and $\Sigma_2$, is the Hilbert-Schmidt norm: $\|\Sigma_1-\Sigma_2\|_{S_2}^2=\sum_{v=1}^{\infty}\lambda_v^2$, where $\{\lambda_1, \lambda_2, \ldots\}$ is the sequence of eigen-values of $\Sigma_1-\Sigma_2$. The clustering objective is to maximize the distance between the two estimated covariance operators. With the assumption that the two clusters are of the same size, they developed a heuristic algorithm that, in each iteration, randomly exchanges members between the two clusters. The problem in Kashlak et al. \cite{Kashlak2019214} is to partition $n$ estimated covariance operators $\{\hat{\Sigma}_i: i=1, \ldots, n\}$ into $K$ clusters, where each $\hat{\Sigma}_i$ is calculated from a few sample functions. An EM-type algorithm was developed where, given the membership values $\{p_{ik}\}_{i=1}^n$, the covariance operator estimate for the $k$th cluster is $\bar{\Sigma}_k=\frac{\sum_{i=1}^np_{ik}\hat{\Sigma}_i}{\sum_{i=1}^np_{ik}}$, and given the covariance operator estimates $\{\bar{\Sigma}_k\}_{k=1}^K$, the membership value $p_{ik}:=\Pr(z_i=k)$ is approximated by a quantity proportional to $\exp(-\frac{n_k(\|\hat{\Sigma}_i-\bar{\Sigma}_k\|_{S_p}-\|R_k\|_{S_p})^2}{2\hat{\gamma}_k^2})$, where $n_k$ is the cluster size, $R_k$ is the Rademacher sum, and $\hat{\gamma}_k^2$ is the empirical weak variance.

Another intuitive approach is to define a notion of centrality (including mean, median, mode, etc.) for a set of functions. Given a set of sample functions $\mathcal{C}=\{y_1, \ldots, y_n\}$, Dabo-Niang et al. \cite{DABONIANG20074878} defined the modal curve in the set as $y_{mo}=\arg\max\limits_{y\in\mathcal{C}}\sum_{i=1}^n\kappa(\frac{\mathbbm{d}(y, y_i)}{\theta})$, where $\kappa(\cdot)$ is a kernel function, and $\theta>0$ is the bandwidth parameter. The median curve of the set is $y_{me}=\arg\max\limits_{y\in\mathcal{C}}\sum_{i=1}^n\mathbbm{d}(y, y_i)$. The heterogeneity of the sample functions in the set $\mathcal{C}$ is measured by the distance between $y_{mo}$ and $y_{me}$: $HI(\mathcal{C})=\frac{\mathbbm{d}(y_{mo}, y_{me})}{\mathbbm{d}(y_{mo}, 0)+\mathbbm{d}(y_{me}, 0)}$. They developed a successively splitting algorithm, for which the stopping criterion is that the reduction in heterogeneity, i.e., $HI(\mathcal{C})-\sum_{k=1}^K\frac{n_k}{n}HI(\mathcal{C}_k)$, should be larger than a threshold, where $\{\mathcal{C}_1, \ldots, \mathcal{C}_K\}$ is a partition of $\mathcal{C}$, and $n_k$ is the size of $\mathcal{C}_k$. The density-based clustering technique for tabular data represents each cluster by a salient mode/density-peak in the population; for example, a state-of-the-art density-based clustering method is given by Tobin and Zhang \cite{JT2021ICDM}. With the attempt to apply density-based clustering methods directly to functional data, Ciollaro et al. \cite{Ciollaro20162922} generalized the kernel smoothing technique to functional data and defined another notion of density: $f(y; \theta)=\mbox{E}_Y[\kappa(\theta^{-1}\|y-Y\|^2)]$, where $\kappa(\cdot)$ is a kernel function, and $\theta$ is the bandwidth parameter. They identified conditions under which the modes of the population, determined by the density $f(y; \theta)$, are well-defined and estimable. In the analysis of EEG data, where a scalp was placed with $n$ channels, and $m$ sample functions (smoothed log-periodograms) were extracted from each channel, Chen et al. \cite{Chen2021425} developed an agglomerative algorithm that iteratively merges two clusters with the minimal $L^2$ distance between their median functions. The $n$ channels, each with $m$ functions, form the initial $n$ clusters in the hierarchy. Baragilly et al. \cite{Baragilly202247} defined the notion of spatial rank function: $f\circ y=\mbox{E}_Y[\|y-Y\|^{-1}(y-Y)]$, to measure the centrality of the function $y$ in the distribution of $Y$; if the norm $\|f\circ y\|$ is close to zero, then the function $y$ is close to the spatial median of $Y$. In the $m$th iteration of the algorithm, with a committee $\mathcal{C}$ of $m$ sample functions, the centrality of any sample function $y_i$ ($1\leq i\leq n$) w.r.t. the committee can be characterized by $\mathbbm{d}(y_i, \mathcal{C})=\|\frac{1}{m}\sum_{\eta\in\mathcal{C}}\frac{y_i-\eta}{\|y_i-\eta\|}\|$. In the $(m+1)$st iteration, the algorithm first updates the committee $\mathcal{C}$ to have $m+1$ most central sample functions according to the $\mathbbm{d}(y_i, \mathcal{C})$'s in the $m$th iteration, then calculates $q_{m+1}=\min\{\mathbbm{d}(y_i, \mathcal{C}): y_i\notin\mathcal{C}\}$, and finally updates the $\mathbbm{d}(y_i, \mathcal{C})$'s.  Clusters are identified from the trace plot of the $q_m$'s. The $k$-medoids type algorithm developed by Cuesta-Albertos and Fraiman \cite{CUESTAALBERTOS20074864} depends on the notion of impartial trimmed means. Given $\alpha\in(0, 1)$, the impartial $\alpha$-trimmed mean of $Y$ is $\mu_{\alpha}=\arg\inf_{\mu\in L^2 (\mathscr{T}, \mathbb{R})}\inf_{\tau\in\mathscr{S}_\alpha}\int\|y-\mu\|^2\tau(y)d\mathbb{P}(y)$, where $\mathscr{S}_\alpha$ is the set of $\alpha$-trim functions: $\mathscr{S}_\alpha=\{\tau: L^2 (\mathscr{T}, \mathbb{R})\mapsto[0, 1], \int\tau(y)d\mathbb{P}(y)\geq 1-\alpha\}$. For the clustering problem, then the objective function is
$\inf_{\mu_1, \ldots, \mu_K\in L^2 (\mathscr{T}, \mathbb{R})}\inf_{\tau\in\mathscr{S}_\alpha}\int(\min_{k=1, \ldots, K}\|y-\mu_k\|^2)\tau(y)d\mathbb{P}(y)$.  The empirical version is obtained by replacing the probability distribution $\mathbb{P}$ with the empirical distribution. To reduce the computational load, they replaced the optimization over the Hilbert space, i.e.,  $\inf_{\mu_1, \ldots, \mu_K\in L^2 (\mathscr{T}, \mathbb{R})}$, with the optimization over the collected sample functions, i.e., $\inf_{\mu_1, \ldots, \mu_K\in \{y_i\}_{i=1}^n}$. Lalo\"{e} \cite{LALOE202151} generalized the traditional $k$-medoids algorithm to functional data. However, the iterative algorithm updates one cluster at a time.

The notion of subspace was also adopted for representing clusters of functions. In  Chiou and Li \cite{Chiou2007} and  Chiou and Li \cite{ChiouLi20082090}, the $k$th subspace is spanned by the mean function $\mu_k$ and the $m_k$ eigen-functions $\{b_{kv}\}_{v=1}^{m_k}$ of the covariance operator $\Sigma_k$. For any sample function $y_i$, its projection onto the $k$th subspace has the form $\mathscr{P}_k(y_i)=\mu_k+\sum_{v=1}^{m_k} a_{ikv}b_{kv}$ in \cite{Chiou2007}, and the form $\mathscr{P}_k(y_i)=\theta_{ik}[\mu_k+\sum_{v=1}^{m_k} a_{ikv}b_{kv}]$ in \cite{ChiouLi20082090}, where the multiplicative factor $\theta_{ik} (>0)$ is a random effect with E($\theta_{ik}$)=1. Then the sample function $y_i$ will be assigned to the $k$th cluster if $\|y_i-\mathscr{P}_k(y_i)\|=\min\{\|y_i-\mathscr{P}_r(y_i)\|: r=1, \ldots, K\}$. The proposed iterative algorithm repeatedly updates the cluster subspaces and the cluster assignment. Bahadori et al. \cite{pmlr-v37-bahadori15} and Guo et al. \cite{Guo2021777} (to be explained in Section \ref{CurveRegistration}) generalized the tabular-data subspace clustering technique to functional data. Bahadori et al. \cite{pmlr-v37-bahadori15} assumed that each cluster is represented by a different manifold, and the self-expressive assumption in the subspace clustering technique indicates that $y_i$ with $z_i=k$ can be expressed as a linear combination of deformed cluster members: $\{g_j(y_j): z_j=k, y_j\neq y_i\}$, where the $g_j$'s are deformations.  By minimizing over $\{a_{ij}, g_j\}_{j=1}^n$ the difference $\|y_i-\sum_{j=1}^na_{ij}g_j(y_j)\|$, subject to certain constraints, we obtain the coefficient matrix $\pmb{A}=[a_{ij}]_{n\times n}$, having zero diagonal entries. The traditional spectral clustering method was applied on the affinity matrix $|\pmb{A}|+|\pmb{A}^T|$.

The Dirichlet process modeling technique is a natural choice for functional data. Petrone et al. \cite{Petrone2009755}, Nguyen \cite{Nguyen2010817} and Nguyen and Gelfand \cite{Nguyen20111249} assumed that the sampling schemes $\{\underline{t}_i\}_{i=1}^n$ are identical with $\underline{t}=\{t_1, \cdots, t_r\}$ and applied the nonparametric Bayesian framework for both the global clustering of the sample functions $\{y_1, \cdots, y_n\}$ and the local clustering of the observations $\{y_1(t_j), \ldots, y_n(t_j)\}$ for every location $t_j\in\underline{t}$. For the global clustering task, the $y_i$'s are generated from a random distribution $G$ that concentrates a probability mass $\pi_k$ on the atomic function $\mu_k$ ($k=1, \ldots, K$): $G=\sum_{k=1}^K\pi_k\delta_{\mu_k}$. The atomic functions $\{\mu_k\}_{k=1}^K$ are i.i.d. according to a probability measure $G^0$. When $K$ is finite, the $\pi_k$'s are from a Dirichlet distribution. When $K$ is infinite, the $\pi_k$'s are given by a stick-breaking process in which the beta distribution is Beta(1, $\theta$); that is, the $\mu_k$'s are generated from a Dirichlet process with the concentration parameter $\theta$ and the base distribution $G^0$. For the local clustering problem, the hidden cluster label $z_i$ may change with $t\in\mathscr{T}$, and the value of $y_i(\underline{t})$ may not be $\mu_k(\underline{t})$, but of the form $(\mu_{k_1}(t_1), \mu_{k_2}(t_2), \ldots, \mu_{k_r}(t_r))^T$, where $k_j\in \{1, \ldots, K\}$ for $j=1, \ldots, r$. Petrone et al. \cite{Petrone2009755} assumed that the atomic functions $\{\mu_k\}_{k=1}^K$ are i.i.d. according to $G^0$, yet the distribution $G$ for the vector $y_i(\underline{t})$ concentrates a probability mass $\pi_{k_1, \ldots, k_r}$ on the mutation $(\mu_{k_1}(t_1), \mu_{k_2}(t_2), \ldots, \mu_{k_r}(t_r))^T$. The $\pi_{k_1, \ldots, k_r}$ has either a Dirichlet prior for finite-mixture modeling, or a stick-breaking prior for infinite-mixture modeling. Nguyen and Gelfand \cite{Nguyen20111249} assumed that the atomic functions $\{\mu_k\}_{k=1}^K$ are from a Gaussian process, and provided some properties of the local clustering model when $K$ is finite. In Nguyen \cite{Nguyen2010817}, the motivating example for local clustering is that the function identity information is not available at any time point $t_j\in\underline{t}$; that is, the data are in the form of $\{y_{\iota_1}(t_j), \ldots, y_{\iota_n}(t_j)\}$, where $\{\iota_1, \ldots, \iota_n\}$ is a random permutation of $\{1, \ldots, n\}$. They developed a hierarchical Dirichlet process model: the distribution of $\{y_{\iota_1}(t_j), \ldots, y_{\iota_n}(t_j)\}$, denoted by $G_j$, is from a Dirichlet process with a concentration parameter $\theta_j$ and a base distribution $G^0_j$. The base distributions $\{G^0_j\}_{j=1}^r$ are conditionally independent draws from another Dirichlet process.

A natural generalization of the traditional Gaussian mixture model to functional data is the Gaussian-process mixture model, where each random function $Y_k$ is a Gaussian process, and $\Pr(z_i=k)=\pi_k$; see, e.g., \cite{NIPS2000_9fdb62f9}, \cite{NIPS2001_9afefc52} and \cite{doi.org/10.1002/acs.744}. A formal definition of Gaussian-process mixture model is given by \cite{SaremSeitz}. To make the residual term in the truncated Karhunen-Lo\`{e}ve expansion have a Gaussian distribution, Zhong et al. \cite{Zhong2021852} applied a nonparametric transformation on the sample functions: given $z_i=k$, $g(y_i(t))=\mu_k(t)+\sum_{v=1}^ma_{ikv} b_{kv}(t)+e_{ik}(t)$, such that $e_{ik}(t)$ is an independent Gaussian white process with variance $\gamma_k^2$. The unknown parameters include $\{\pi_k, \gamma_k^2\}_{k=1}^K$  and the nonparametric transformation function $g$. An EM-type algorithm was developed for parameter estimation. A follow-up work \cite{cjs.11680} generalizes the setting to include functional predictors. Given the sampling scheme $\underline{t}_i$, the collected data are in the form of $\{(\tilde{y}_i(t_{ir}), \beta_i(t_{ir})): r=1, \ldots, r_i\}$, where $\beta_i=(\beta_i^1, \ldots, \beta_i^q)^T$ are $q (\geq 1)$ functional predictors. Given $z_i=k$, the regression model is $g(y_i(t))=\mu_k(t)+\sum_{j=1}^q\alpha_{kj} \beta_i^j(t)+e_{ik}(t)$, where $e_{ik}(t)$ is an independent Gaussian process with zero mean and a parametric kernel function $\Sigma_k(\cdot, \cdot)$. Each unknown function $\mu_k$ is approximated by a set of B-spline basis functions.

The idea in H\'{e}brail et al. \cite{HEBRAIL20101125} and Chamroukhi \cite{Chamroukhi2016374} is to represent each cluster by a simple function, e.g., piecewise linear/polynomial regression models. In \cite{HEBRAIL20101125}, each cluster is represented by a piecewise constant/linear function. With the user specifying the number of segments for each cluster prototype (i.e., the piecewise constant/linear function), the developed $k$-means type algorithm iteratively updates the cluster prototypes and the cluster assignment (according to the $L^2$ distance). With the assumption that the sampling schemes $\{\underline{t}_i\}_{i=1}^n$ are identical with $\underline{t}=\{t_1, \ldots, t_r\}$, the segmentation problem of the interval $\mathscr{T}$ reduces to the segmentation of the discrete points $\{t_1, \ldots, t_r\}$, and hence the cluster prototypes can be readily calculated by dynamic programming. In \cite{Chamroukhi2016374}, the piecewise polynomial formulation of the $k$th mean function is $\mu_k(t)=\sum_{v=1}^{m_k}g(t; \pmb{\theta}_{kv})\delta(t\in \mathscr{T}_{kv})$, where the function domain $\mathscr{T}$ is partitioned into $m_k$ disjoint intervals $\{\mathscr{T}_{k1}, \ldots, \mathscr{T}_{km_k}\}$ with $\mathscr{T}=\cup_{v=1}^{m_k}\mathscr{T}_{kv}$, and $g(\cdot; \pmb{\theta}_{kv})$ is a polynomial regression model with regression coefficient vector $\pmb{\theta}_{kv}$. Given $z_i=k$, the Gaussian white noise $\epsilon_i(t)$ has the variance $\sigma^2_{kv}$ when $t\in \mathscr{T}_{kv}$. An EM-type algorithm was developed to estimate both the unknown model parameters and the segmentation $\{\mathscr{T}_{k1}, \ldots, \mathscr{T}_{km_k}\}$ for each cluster.

Zambom et al. \cite{Zambom2019527} developed a $k$-means type algorithm, where a cluster center is the average of the member functions. However, the allocation of a function to a cluster is decided by two hypothesis tests: the one-sample t-test and the one-way ANOVA test. The motivation is that, if $z_i=k$, then the residuals $\{\tilde{y}_i(t_{ir})-\mu_k(t_{ir}): r=1, \ldots, r_i\}$ will have a constant mean. The value of the t-test statistic can be readily calculated from the $r_i$ residuals. The ANOVA test statistic requires each group to have three or more data points. For every sample function $y_i$, they created $r_i$ groups, with the $r$th group having $m$ data points (the residual $\tilde{y}_i(t_{ir})-\mu_k(t_{ir})$ and its $m-1$ neighboring residuals). Then the within-groups sum-of-squares and between-groups sum-of-squares were calculated in the traditional way. A follow-up work in Zambom et al. \cite{Zambom2022} extends the hypothesis testing approach to determine the number of clusters. The clustering methods in \cite{Zambom2019527} and \cite{Zambom2022} both have strong assumptions attached to them and hence are of limited application.

\begin{figure}[!ht]
	\centering
	\includegraphics[width=14cm]{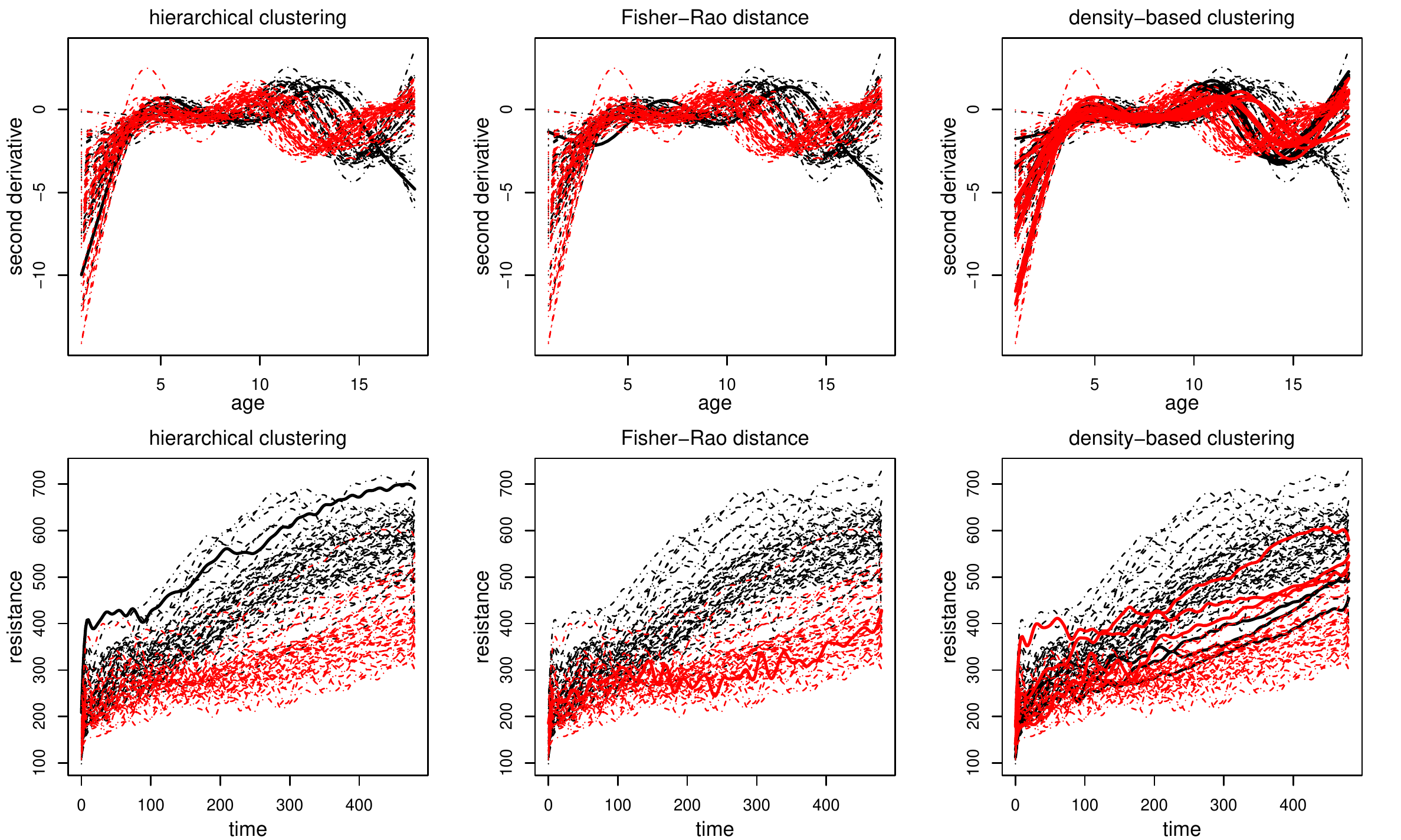}
	\caption{Left: The agglomerative hierarchical clustering technique with the $L^2$ distance. Middle: The agglomerative hierarchical clustering technique with the Fisher-Rao distance. Right: The density-based clustering technique with the $L^2$ distance. The upper panels correspond to the growth acceleration curves, and the lower panels correspond to the dough resistance curves. Incorrectly clustered curves are indicated by bold solid lines. For example, a red bold solid line in the top left panel means that a girl's curve was assigned into the boy's cluster. The hierarchical clustering technique (with an appropriate linkage method) outperforms all the other methods, including the three methods in Figure \ref{example_VS}.}
\label{example_FS}
\end{figure}

In Figure \ref{example_FS}, 
we compare two different clustering methods and two different distance metrics, the $L^2$ distance metric and the Fisher-Rao distance metric. The Fisher-Rao distance metric has the property that it is invariant to homeomorphism time-warping functions \cite{arxiv.1103.3817}. The left and middle panels differ in the distance metric, while the left and right panels differ in the clustering method. For the hierarchical clustering technique, after trying all the linkage methods available in the R function ``hclust'', we adopted the centroid linkage method for the growth acceleration curves and the single linkage method for the dough resistance curves. The density-based clustering method is from the R package \textit{FADPclust}. We here do not implement the functional $k$-means algorithm with the $L^2$ distance metric, because it is equivalent to the combination of the multivariate $k$-means algorithm and the fPC decomposition (implemented in Figure \ref{example_VS}). We visualize the clustering results through the estimated smooth functions, where a curve in the form of bold solid line means that it is incorrectly clustered.

\section{Vector-Valued Functional Data}\label{Vector-Valued}
Vector-valued functional data are also frequently referred to as multivariate/multi-dimensional functional data. Cluster analysis of vector-valued functional data is much more challenging, mainly due to the complex dependency among the component random functions. A few recent clustering methods for multivariate time series can be found in \cite{1.97816119743}, \cite{3097983.3098060}, \cite{pmlr-v84-saad18a} and \cite{978-3-030-47}.

The Karhunen-Lo\`{e}ve expansion of the matrix of variance-covariance functions, $\Sigma(s, t)=\mbox{E}[(Y(s)-\mu(s))(Y(t)-\mu(t))^T]$, has the form
\begin{equation*}
\Sigma(s, t)=\sum_{v=1}^{\infty}\lambda_v b_v(s)b_v^T(t),
\end{equation*}
where the $b_v$'s are orthonormal vector-valued eigen-functions. Then the fPC decomposition of $y_i$ (a realization of $Y$) admits the form: $y_i(t)=\mu(t)+\sum_{v=1}^{\infty}a_{iv} b_v(t)$, where $a_{iv}=\langle y_i-\mu, b_v\rangle$ is the fPC score associated with the eigen-function $b_v$, with the properties E$[a_{iv}]=0$, var$(a_{iv})=\lambda_v$, and E$[a_{iv}a_{ir}]=0$ for $v\neq r$. Note that, for vector-valued functional data, it is common to standardize each component function before performing any statistical analysis, to account for differences in degrees of variability and in units of measurements among the component random functions. The standardization of functional data is analogous to that of tabular data; that is, each component function $y_i^j$ is subtracted by its mean and then divided by its standard deviation: $\frac{y_i^j(t)-\mbox{E}[Y^j(t)]}{\sqrt{\mbox{var}(Y^j(t))}}$, for $j=1, \ldots, p$.

\subsection{Clustering in Finite-Dimensional Space}
Basis expansion of vector-valued functions is commonly achieved by the eigen-functions of the covariance operator. Jacques and Preda \cite{JACQUES201492} applied the model in \cite{JACQUES2013164} to vector-valued functional data: given $\Sigma_k(s, t)=\sum_{v=1}^{\infty}\lambda_{kv} b_{kv}(s)b_{kv}^T(t)$, $z_i=k$, and $y_i(t)=\mu_k(t)+\sum_{v=1}^{m_k}a_{ikv} b_{kv}(t)$, they assumed that the distribution of the fPC-score vector $(a_{ik1}, \ldots, a_{ikm_k})^T$ is $\mathscr{N}(\pmb{0}, \mbox{diag}(\lambda_{k1}, \ldots, \lambda_{km_k}))$. Schmutz et al. \cite{Schmutz20201101} assumed that the distribution of the fPC-score vector is $\mathscr{N}(\pmb{u}_k, \mbox{diag}(\gamma_{k1}$, $\ldots, \gamma_{km_k}, \gamma_k, \ldots, \gamma_k))$. For longitudinal functional data with missing values, Bruckers et al. \cite{WOS:000408988700020} took the ensemble clustering approach and applied the Gaussian-mixture model in \cite{JACQUES201492} on each of a few randomly imputed functional datasets to obtain the base clusterings. The final clustering, i.e. the consensus clustering, is the median partition of the base clusterings \cite{Zhang2022PR}. In Golovkine et al. \cite{GOLOVKINE2022107376}, a sample function $y_i$ has the form: $y_i(t)=\sum_{k=1}^K\mu_k(t)\delta(z_i=k)+\sum_{v=1}^{\infty}\alpha_{iv}\beta_v(t)$, where $\{\beta_v\}_{v=1}^\infty$ is an orthonormal basis of $L^2(\mathscr{T}, \mathbb{R}^p)$, and $[\alpha_{iv}|z_i=k]$ has an independent Gaussian distribution $\mathscr{N}(0, \gamma_{kv}^2)$. Let $\{b_v\}_{v=1}^\infty$ be the orthonormal eigen-functions from the Karhunen-Lo\`{e}ve expansion of the population covariance function $\Sigma(s, t)=\sum_{k=1}^K\pi_k\Sigma_k(s, t)$. Define $\mu=\sum_{k=1}^K\pi_k\mu_k$ and write $y_i=\mu(t)+\sum_{v=1}^{\infty}a_{iv} b_v(t)$, where $a_{iv}=\langle y_i-\mu, b_v\rangle$. Then, given $z_i=k$, the distribution of $a_{iv}$ is $\mathscr{N}(\langle \mu_k-\mu, b_v\rangle, \sum_{l=1}^{\infty}\gamma_{kl}^2\langle \beta_l, b_v\rangle^2)$, and cov($a_{iv}, a_{ij}$)=$\sum_{l=1}^{\infty}\gamma_{kl}^2\langle \beta_l, b_v\rangle\langle \beta_l, b_j\rangle\neq0$. Therefore, given $z_i=k$, they assumed that the fPC-score vector $\pmb{a}_i$ has a general Gaussian distribution. The proposed clustering algorithm repeatedly splits a dataset into two by performing the fPC decomposition on the current data and then applying the traditional Gaussian mixture model (with only two mixture components) on the fPC-score vectors.

Other basis systems and/or classical clustering methods were also adopted. Kayano et al. \cite{Kayano2010211} applied orthonormalized Gaussian basis functions for the smoothing of each component sample path. The $p$ component coefficient vectors were concatenated into one large vector, and the self-organizing map method was applied to cluster the $n$ concatenated coefficient vectors. Serban and Jiang \cite{Serban2012805} adopted the functional analysis of variance model: for $j=1, \ldots, p$, $y_i^j=\chi_i+x_i^j$, where the ``mean function'' $\chi_i$ is shared by all component sample functions $\{y_i^1, \ldots, y_i^p\}$. They performed two levels of clustering, one according to the mean functions only $\{\chi_1, \ldots, \chi_n\}$, and the other according to the vector-valued deviation functions only $\{x_1, \ldots, x_n\}$. By applying a nonparametric decomposition: $y_i^j(t)=\sum_{v=1}^{m_1}\alpha_{i v} \beta_v(t)+\sum_{v=1}^{m_2}a_{iv}^j b_v(t)$, where both $\{\beta_v\}_{v=1}^\infty$ and $\{b_v\}_{v=1}^\infty$ are orthogonal, traditional clustering methods were applied on the coefficient vectors $\{\pmb{\alpha}_i=(\alpha_{i 1}, \ldots, \alpha_{i m_1})^T\}_{i=1}^n$ for level-one clustering, and on the concatenated coefficient vectors $\{\pmb{a}_i=(a_{i1}^1, \ldots, a_{im_2}^1, \ldots, a_{i1}^p, \ldots, a_{im_2}^p)^T\}_{i=1}^n$ for level-two clustering.

Ben Slimen et al. \cite{BENSLIMEN201897} arranged the vector-valued sample functions in a matrix, where the component sample function $y_i^j$ ($1\leq i\leq n$, $1\leq j\leq p$) is in the $i$th row and $j$th column. They performed co-clustering that divides the sample function matrix into $K_r$ row-clusters and $K_c$ column-clusters, i.e., $K_r\times K_c$ blocks, where the row and column cluster-indicator variables are independent. By projecting every component sample function onto the same linear space, $y_i^j(t)=\sum_{v=1}^ma_{iv}^j b_v(t)$, they assumed that the coefficient vectors $\{\pmb{a}_i^j: i=1, \ldots, n, j=1, \ldots, p\}$ have a Gaussian mixture model (with $K_r\times K_c$ Gaussian components). The co-clustering framework was again adopted by Ben Slimen et al. \cite{BenSlimen20}, where each subject is characterized by both a vector-valued random function and multiple binary variables. The matrix is divided into $K_r$ row-clusters and $K_c=K_f+K_b$ column-clusters; that is, component random functions are partitioned into $K_f$ clusters, and binary variables into $K_b$ clusters. Within each block, the component random functions and the binary variables are independent, and the binary data are modeled by the Bernoulli distribution. A functional co-clustering method is given by Galvani et al. \cite{Galvani2021}, therein called bi-clustering. Let $(I, J)$ be the index set for, e.g., the $k$th cluster, where $I$ (resp. $J$) denote a subset of rows (resp. columns). Let $|I|$ denote the size of the set $I$. The within-cluster variation is defined as $\frac{1}{|I| |J|}\sum_{i\in I, j\in J}\|y_{ij}-(\mu_k+x_{ki\cdot}+x_{k\cdot j})\|^2$, with the estimates $\mu_k=\frac{1}{|I| |J|}\sum_{i\in I, j\in J}y_{ij}$, $x_{ki\cdot}=\frac{1}{|J|}\sum_{j\in J}(y_{ij}-\mu_k)$ and $x_{k\cdot j}=\frac{1}{|I|}\sum_{i\in I}(y_{ij}-\mu_k)$. The algorithm repeatedly removes and adds indices to $I$ and $J$ to find the biggest cluster with the within-cluster variation below a threshold value.
In Bouveyron et al. \cite{bouveyron:hal-02862177}, each 6-year long sample function $y_i$ (i.e., a time series) was cut into 313 one-week long sample functions $\{y_{ir}: r=1, \ldots, 313\}$; that is, the time domain changes to one week. They arranged the sample functions into a matrix and developed a co-clustering method to identify homogeneous blocks of geographical locations (rows) and weeks (columns). The proxy for each vector-valued sample function $y_{ir}$ is the concatenated coefficient vector $\pmb{a}_{ir}=(a_{ir1}^1, \ldots, a_{irm}^1, \ldots, a_{ir1}^p, \ldots, a_{irm}^p)^T$ with $y_{ir}^j(t)=\sum_{v=1}^m a_{irv}^j b_v(t)$. By assuming that the row and column cluster-indicator variables are independent, the subspace model in Bouveyron and Jacques  \cite{Bouveyron2011281} was applied to build a Gaussian mixture model.

\subsection{Clustering in Infinite-Dimensional Space}
Except for the co-clustering work \cite{Galvani2021} explained above, all the other articles in this category are focused on defining a (dis)similarity measure for vector-valued functional data. Tokushige et al. \cite{Tokushige20071}, Ieva et al. \cite{Ieva2013401}, Meng et al. \cite{Meng2018166} and Martino et al. \cite{Martino2019301} all applied the $k$-means algorithmic framework; the distance metric in \cite{Tokushige20071} is $\mathbbm{d}_0$, and in both \cite{Ieva2013401} and \cite{Meng2018166} is $\sqrt{\mathbbm{d}_0^2+\mathbbm{d}_1^2}$. In \cite{Martino2019301}, the squared distance between $y_i$ and $y_j$ is $\int_{0}^\infty\sum_{v=1}^{\infty}\langle y_i-y_j, b_v\rangle^2\exp(-\lambda_vw)g(w)dw$, where the eigen-components $\lambda_v$ and $b_v$ are related to the population covariance operator, and the non-increasing function $g$ is to make the integration converge. The following two works, \cite{Bruno2011975} and \cite{Li201615}, applied the $k$-medoids algorithmic framework. The functional data in Bruno et al. \cite{Bruno2011975} are sequences of composition measurements, and therefore the measurements $\{y^1_i(t), \ldots, y^p_i(t)\}$ at any time $t$  sum up to one. They calculated the Aitchison distance between $y_i$ and $y_j$, averaged over a common grid. The pairwise distance matrix is the input to the $k$-medoids algorithm. In Li et al. \cite{Li201615}, the two component random functions of $Y=(Y^1, Y^2)$ are water temperature and air temperature stochastic processes that have the relation: $y_i^1(t)=a_{1i}(t)+a_{2i}(t)y_i^2(t)+e_i(t)$, where $a_{1i}$ and $a_{2i}$ are respectively the function intercept and function slope, and $e_i$ is the error term. The distance between two sample functions $y_i$ and $y_j$ is $g(\|\pmb{s}_i-\pmb{s}_j\|_2)[w\mathbbm{d}_c(a_{1i}, a_{1j})+(1-w)\mathbbm{d}_c(a_{2i}, a_{2j})]$, where $\|\pmb{s}_i-\pmb{s}_j\|_2$ is the Euclidean distance between the geographical location of $y_i$ and that of $y_j$, and $\mathbbm{d}_c$ is the Canberra distance. Then the traditional $k$-medoids algorithm was applied. Chen et al. \cite{doi2018.1494280} coupled the self-organizing map method with either $\mathbbm{d}_0$ or $\mathbbm{d}_1$. In particular, by initializing $c_i^0$ with $y_i$, for $i=1, \ldots, n$, the algorithm repeatedly updates $\{c_1^v, \ldots, c_n^v\}$ via $c_i^v=\frac{\sum_{j=1}^n\kappa(c_j^{v-1}, c_i^{v-1})c_j^{v-1}}{\sum_{j=1}^n\kappa(c_j^{v-1}, c_i^{v-1})}$, until either $\max\{\mathbbm{d}_0(c_i^{v-1}, c_i^v): i=1, \ldots, n\}$ or $\max\{\mathbbm{d}_1(c_i^{v-1}, c_i^v): i=1, \ldots, n\}$ is smaller than a threshold. The function $\kappa$ is the Gaussian kernel with either the $\mathbbm{d}_0$ metric or the $\mathbbm{d}_1$ semi-metric.

\begin{figure}[!ht]
	\centering
	\includegraphics[width=14cm]{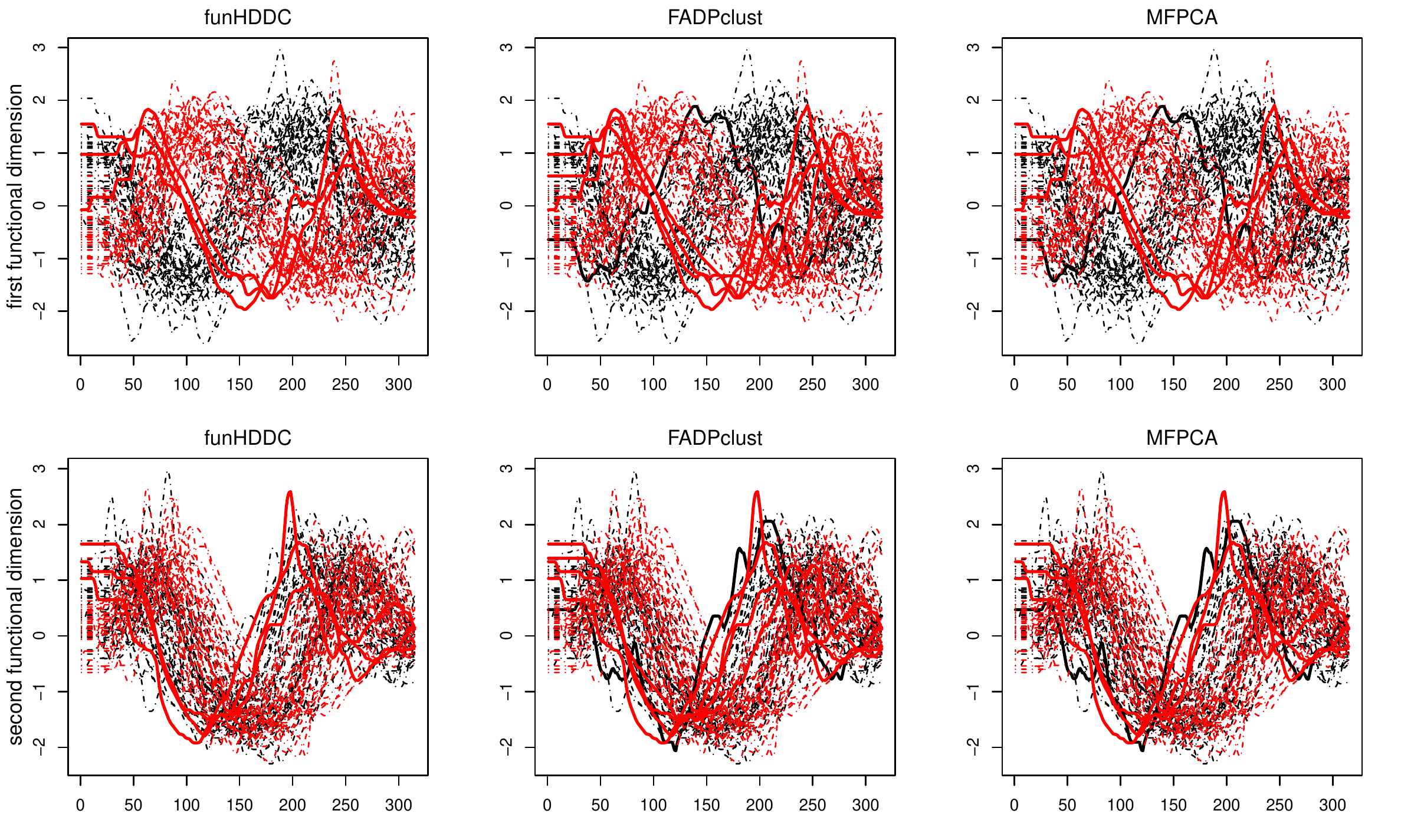}
	\caption{Clustering results of vector-valued functional data ($p=2$ and $K=2$). The upper panels correspond to the X coordinate, and the lower panels correspond to the Z coordinate. Incorrectly clustered curves are indicated by bold solid lines. The curves on the X coordinate have a clear clustering pattern, while the curves on the Z coordinate are difficult to partition. Hence, the good clustering results here are mainly because the curves on the X coordinate have two distinct shapes.}
\label{example_multi-D}
\end{figure}

In Figure \ref{example_multi-D}, 
we apply three different clustering methods on a two-dimensional functional dataset (i.e., $p=2$), where the functions are from two classes. The data are provided by \cite{4912759} and available at the UEA \& UCR Time Series Classification Repository (i.e., the ``UWaveGestureLibrary'' dataset). The original data consist of over 4000 instances, each of which is a sequence of accelerometer readings of one gesture in three dimensions (i.e.  the X, Y, Z coordinates). There are eight different gestures/classes. The data in Figure \ref{example_multi-D} are from the testing dataset. The two dimensions correspond to the X and Z coordinates, and the two classes correspond to the 7th and 8th gestures in Figure 3 of \cite{4912759}. The first clustering method is built on the work \cite{Schmutz20201101}, available in the R package \textit{funHDDC}. The second clustering method is the density-based clustering method in the R package \textit{FADPclust}. The third clustering method is a combination of the multivariate fPC analysis and the $k$-means algorithm; that is, applying the multivariate $k$-means algorithm on the fPC score vectors. The R package \textit{MFPCA} was used for obtaining the fPC scores.

\section{Dependent Functional Data}\label{Dependent}
A ubiquitous type of dependent functional data is called the spatio-temporal data. There exists a vast literature on approaches for clustering spatio-temporal data. Two recent review articles are \cite{10.1145/3161602} and \cite{Ansari20202381}, with the former covering both supervised and unsupervised data-mining techniques for spatio-temporal data.  Let $\mathscr{T}=\mathscr{T}_s\times\mathscr{T}_t$ denote a spatio-temporal continuum, where $\mathscr{T}_s$ is a spatial region, and $\mathscr{T}_t$ is a time interval. Now the random function $Y$ is a family of random variables $Y=\{Y(\pmb{s}, t): \pmb{s}\in\mathscr{T}_s, t\in\mathscr{T}_t\}$, each defined on $(\Omega, \mathscr{F}, \mathbb{P})$ and taking values in $\mathbb{R}^p$ ($p\geq1$). By fixing the spatial parameter $\pmb{s}$ at a few locations $\{\pmb{s}_1, \ldots, \pmb{s}_n\}$, we often write $y_i(t)=y(\pmb{s}_i, t)$; that is, the sample function $y_i$ is a random sample of $Y(\pmb{s}_i, \cdot)$. The typical clustering problem is to partition the sample functions $\{y_i\}_{i=1}^n$ into groups of different morphological patterns, with the objective to obtain a partition of the spatial locations $\{\pmb{s}_i\}_{i=1}^n$, or even the spatial region $\mathscr{T}_s$. Alternatively, the clustering objective could be to partition the time stamps $\{t_1, \ldots, t_r\}$, not the spatial locations $\{\pmb{s}_1, \ldots, \pmb{s}_n\}$. For example, in the application of discovering patterns of human brain activity from fMRI data, the goal is to identify the time points at which similar brain activity is observed in the brain. Then it is more convenient to fix the time parameter $t$ and treat $\{Y(\cdot, t_1), \ldots, Y(\cdot, t_r)\}$ as a set of time-indexed random fields. Each sample function $y(\cdot, t_v)$ ($1\leq v\leq r$) becomes a spatial map, and the set of evaluations $\{y(\pmb{s}_i, t_v)\}_{i=1}^n$  are to be regarded as one functional datum. The clustering objective is to find groups of time stamps that, within each group, the spatial maps are alike.

\subsection{Clustering in Finite-Dimensional Space}
Secchi et al. \cite{SECCHI201353} developed an ensemble clustering algorithm that generates every base clustering in three steps: (1) randomly sample a set of $m$ nuclei $\{\ddot{\pmb{s}}_1, \ldots, \ddot{\pmb{s}}_m\}$ from the $n$ locations $\{\pmb{s}_1, \ldots, \pmb{s}_n\}$, and obtain the Voronoi tessellation $\{V(\ddot{\pmb{s}}_v)\}_{v=1}^m$ according to a given distance metric $\mathbbm{d}$; (2) calculate the representative function for each cell of the tessellation: for $v=1, \ldots, m$, $\breve{y}_v=\frac{\sum_{\pmb{s}_i\in V(\ddot{\pmb{s}}_v)}\kappa(\mathbbm{d}(\pmb{s}_i, \ddot{\pmb{s}}_v))y_i}{\sum_{\pmb{s}_i\in V(\ddot{\pmb{s}}_v)}\kappa(\mathbbm{d}(\pmb{s}_i, \ddot{\pmb{s}}_v))}$, where $\kappa(\cdot)$ is a kernel function; (3) perform basis expansion on the $m$ representative functions and then partition the $m$ coefficient vectors into $K$ clusters. Abramowicz et al. \cite{Abramowicz2019} developed a two-step clustering framework. In the first-step, a simple clustering method for functional data, e.g., the functional $k$-means algorithm with the $L^2$ distance, was applied to obtain $K_1$ clusters. In the second-step, following \cite{SECCHI201353}, the region $\mathscr{T}_s$ was randomly partitioned into $m$ Voronoi cells $\{V(\ddot{\pmb{s}}_v)\}_{v=1}^m$. However, the representative of each cell is the vector of relative frequencies of the $K_1$ cluster labels in that cell, and a base clustering with $K_2$ clusters is obtained by applying a multivariate clustering method on the relative-frequency vectors.

White and Gelfand \cite{White2021586} applied the mixed-effects model for vector-valued functional data: for any $1\leq j\leq p$, the decomposition is $y_i^j(t)=\mu_i^j(t)+x_i^j(t)=\sum_{v=1}^{q}a^j_{iv} b_v(t)+\sum_{v=q+1}^{m}a^j_{iv} b_v(t)+x_i^j(t)$; that is, the set of basis functions is identical for all component random functions. The $x_i^j$'s are zero-mean Gaussian processes that are dependent across $i$ and across $j$. For any $1\leq j\leq p$, only the coefficient vector $\pmb{a}^j_i=(a^j_{i1}, \ldots, a^j_{iq})^T$ is used for clustering. In particular, for the component-wise clustering task, they assumed that the coefficient vectors $\{\pmb{a}^j_i\}_{i=1}^n$ are draws from a Dirichlet process. With the fPC decomposition $y_i(t)=\mu(t)+\sum_{v=1}^{m}a_{iv} b_v(t)$, Margaritella et al. \cite{Margaritella2021167} applied the Dirichlet process model on the fPC scores $\{a_{1v}, \ldots, a_{nv}\}$ independently for each eigen-dimension $v=1, \ldots, m$. In particular, the fPC scores $\{a_{1v}, \ldots, a_{nv}\}$ are independently generated from a random distribution $G_v$ that concentrates a probability mass $\pi_{kv}$ on the atomic Gaussian distribution $\mathscr{N}(u_{kv}, \gamma_{kv})$. The prior on $u_{kv}$ is $\mathscr{N}(0, \xi_v)$, and on $\gamma_{kv}$ is the gamma distribution $\Gamma(1, \beta_v)$. The $\pi_{kv}$'s are given by a stick-breaking process in which the beta distribution Beta($1, \theta_v$) has a uniform prior on $\theta_v$.

Liang et al. \cite{Liang2021116} also adopted the mixed-effects model: given $z_i=k$, the decomposition is $y_i=\mu_k+x_{ik}$, where the $x_{ik}$'s are spatially correlated random effects. Given the fPC decomposition: $x_{ik}(t)=\sum_{v=1}^{\infty}a_{ikv} b_{kv}(t)$, they assumed that the distribution of the fPC-score vector $(a_{ik1}, \ldots, a_{ikm})^T$ is $\mathscr{N}(\pmb{0}, \pmb{\Gamma}_k)$, where the covariance between the fPC scores $a_{ikv}$ and $a_{jor}$ is cov($a_{ikv}, a_{jor}$)=$\gamma^2_{kv}g_{kv}(\pmb{s}_i-\pmb{s}_j)\delta(k=o, v=r)$, and $g_{kv}$ is a spatial correlation function. A locally dependent Markov random field model was employed for the cluster membership to account for spatial dependence: $\Pr(z_i=k)=\frac{\exp(\theta\sum_{\pmb{s}_j\in \mathscr{S}_i}\delta(z_j=k))}{\sum_{k=1}^K\exp(\theta\sum_{\pmb{s}_j\in \mathscr{S}_i}\delta(z_j=k))}$, where $\mathscr{S}_i$ is the set of neighboring locations of $\pmb{s}_i$.

\subsection{Clustering in Infinite-Dimensional Space}
The typical approach is to incorporate the spatial correlation between two sites $\pmb{s}_i$ and $\pmb{s}_j$ into the dissimilarity measure between $y_i$ and $y_j$. Assuming that the random functions are second-order stationary and isotropic, the dissimilarity measure defined by Giraldo et al. \cite{Giraldo2012403} is $\varpi_{ij}\|y_i-y_j\|$, where the weight $\varpi_{ij}$ is calculated from the trace-variogram function: $\varpi_{ij} = \frac{1}{2}\mbox{E}[\|Y(\pmb{s}_i, \cdot)-Y(\pmb{s}_j, \cdot)\|^2]$. The sample estimate of the expectation $\mbox{E}[\|Y(\pmb{s}_i, \cdot)-Y(\pmb{s}_j, \cdot)\|^2]$ is  $\frac{1}{|\mathscr{S}_{ij}|}\sum_{(\pmb{s}_r, \pmb{s}_v)\in\mathscr{S}_{ij}}\|y_r-y_v\|^2$, where $\mathscr{S}_{ij}=\{(\pmb{s}_r, \pmb{s}_v): -\epsilon\leq\|\pmb{s}_r-\pmb{s}_v\|_2-\|\pmb{s}_i-\pmb{s}_j\|_2\leq \epsilon\}$ for a small $\epsilon$, and $|\mathscr{S}_{ij}|$ is the size of the set $\mathscr{S}_{ij}$. A hierarchical clustering method was applied on the pairwise distance matrix. Utilizing the notion of trace variogram, Romano et al. \cite{Romano2017645} substituted the sample function $y_i(t)$ with $g_i(\vartheta):=\sum_{\pmb{s}_j\in\mathscr{S}_i}\|y_i-y_j\|^2$, where $\mathscr{S}_i=\{\pmb{s}_j: -\epsilon\leq\|\pmb{s}_j-\pmb{s}_i\|_2-\vartheta\leq \epsilon\}$ for a small $\epsilon$. The clustering problem is then to partition the $n$ functions $\{g_i(\vartheta)\}_{i=1}^n$ into $K$ clusters to minimize $\sum_{k=1}^K\sum_{z_i=k}\sum_{\vartheta\in\Theta}[g_i(\vartheta)-\bar{g}_k(\vartheta)]^2$, where $\Theta$ is a set of values for $\vartheta$, and $\bar{g}_k$ is the cluster-wise sample average. Modifying the distance defined by \cite{biom.12161Chen}, i.e., $\|y-\eta\|^2_w=\int_{\mathscr{T}}w(t)[y(t)-\eta(t)]^2dt$, Romano et al. \cite{Romano2020} introduced the trace-variogram function into the coefficient of variation; that is, the weighting function $w$ is obtained by minimizing $\frac{\sum_{i\neq j}\mbox{var}(\|\varpi_{ij}(y_i-y_j)\|^2_w)}{\sum_{i\neq j}\mbox{E}^2[\|\varpi_{ij}(y_i-y_j)\|^2_w]}$. The functional data in Haggarty et al. \cite{Haggarty2015491} are nitrate-concentration time series collected by monitoring stations in a directed river network, and therefore the spatial correlation between two monitoring stations is affected by the flow connectedness. The dissimilarity measure is defined as $g(\mathbbm{d}(\pmb{s}_i, \pmb{s}_j))\|y_i-y_j\|$, where $\mathbbm{d}(\pmb{s}_i, \pmb{s}_j)$ is the stream distance (the shortest distance between two locations along the stream network), and $g(\cdot)$ is the tail-up model in which the covariance function is the Mat\'{e}rn function.

In Shi and Wang \cite{Shi2008267}, each subject is characterized by a sample function $y_i$ and two covariate vectors $\pmb{c}_i$ and $\pmb{\alpha}_i=(\alpha_{i1}, \ldots, \alpha_{iq})^T$. The covariates $\{\pmb{c}_i\}_{i=1}^n$ are related to the heterogeneity in the subjects, and hence are used to model the cluster membership $\Pr(z_i=k)=\frac{\exp(\pmb{c}_i^T\pmb{\beta}_k)}{1+\sum_{v=1}^{K-1}\exp(\pmb{c}_i^T\pmb{\beta}_v)}$. Each $\pmb{\alpha}_i$ is a vector of explanatory variables hence is incorporated in a functional regression model for predicting the functional response $y_i$. The data collected for the sample function $y_i$ are in the form of $\{(t_{ir}, \pmb{s}_{ir}, \tilde{y}_i(t_{ir}, \pmb{s}_{ir})), r=1, \ldots, r_i\}$, where the $\{\pmb{s}_{ir}\}$ are called functional covariates and are modeled by a Gaussian process. In particular, given $z_i=k$, the functional regression model is $y_i(t, \pmb{s})=\sum_{j=1}^q\alpha_{ij} g_{kj}(t)+e_k(\pmb{s})$, where $e_k(\pmb{s})$ is a Gaussian process with zero mean and kernel function $\Sigma_k(\cdot, \cdot)$. Each unknown function $g_{kj}(t)$ is approximated by a set of $m$ basis functions: $g_{kj}(t)=\sum_{v=1}^{m}a_{kjv}b_v(t)$. An EM-type algorithm was developed for parameter estimation. The approach in \cite{Shi2008267} was taken by \cite{9679476}, where each subject is characterized by a sample function $y_i$ and a covariate vector $\pmb{\alpha}_i$. The regression model was simplified to be $y_i(t)=\sum_{j=1}^q\alpha_{ij} g_{kj}(t)$, given $z_i=k$. However, they applied the logistic regression model on the geographical locations: $\Pr(z_i=k)=\frac{\exp(\pmb{s}_i^T\pmb{\beta}_k)}{1+\sum_{v=1}^{K-1}\exp(\pmb{s}_i^T\pmb{\beta}_v)}$, which is impractical.

\begin{figure}[!ht]
	\centering
	\includegraphics[width=15cm]{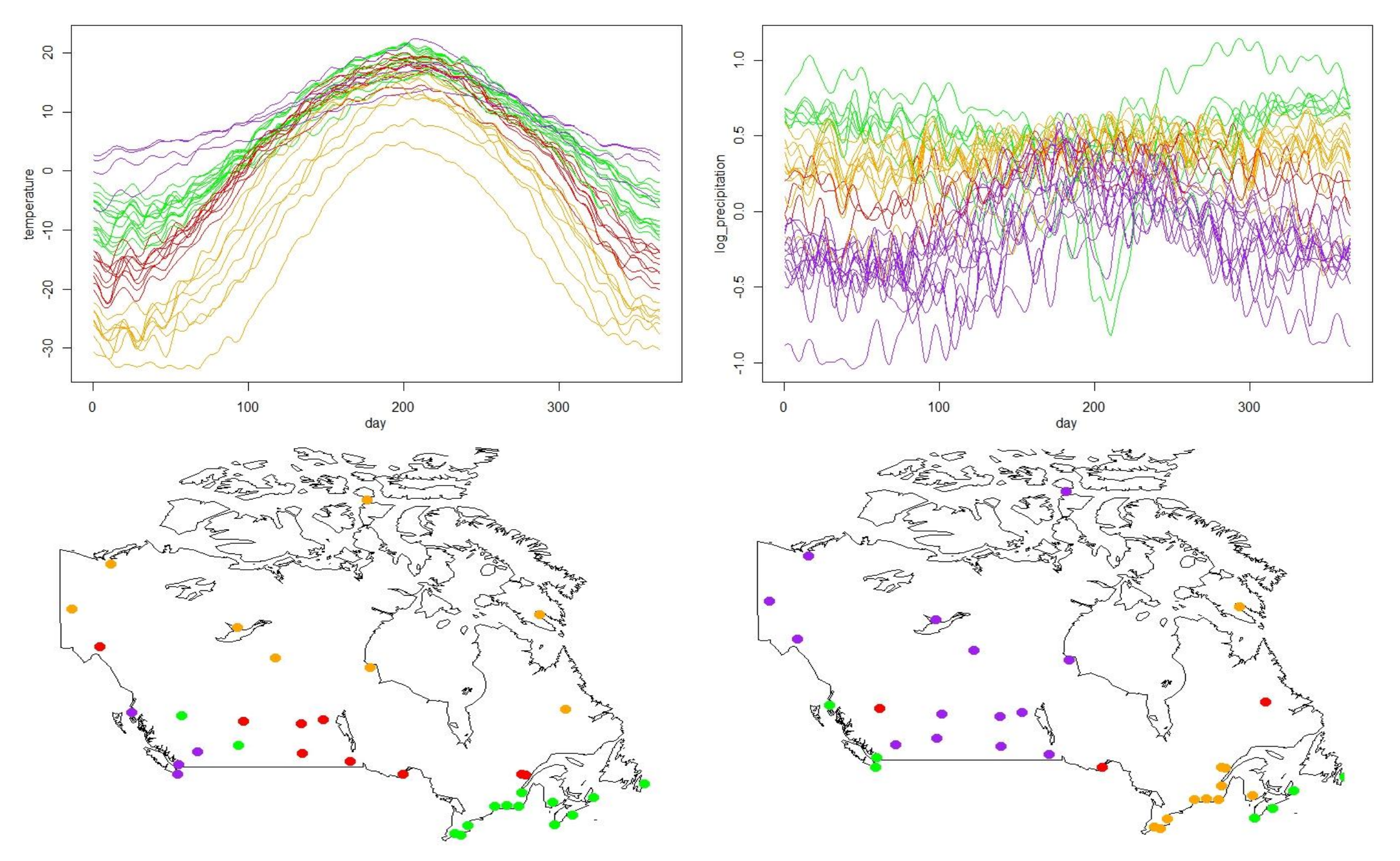}
	\caption{Top left: Clustering results obtained from the temperature data only, where the four colors are orange, red, green and purple. Top right: Clustering results obtained from the log-precipitation data only. The bottom-left panel (resp, bottom-right panel) visualizes the clusters of the temperature data (resp, log-precipitation data) on the Canada map. The upper panels demonstrate that the clusters are of high quality in that curves of similar shapes are in the same cluster. The lower panels imply that, for both datasets, a few curves are incorrectly clustered. Better clustering results could be achieved if temperature and precipitation are used together to define a cluster.}
\label{example_depend}
\end{figure}

We here apply the mixed-effects model developed by \cite{Liang2021116} on the Canadian temperature data, available in the R package \textit{fda}. The dataset consists of daily average temperature and precipitation of 35 Canadian cities for one year. The cities belong to four different regions (i.e., climate zones), and therefore we set the cluster number to be 4. We adopted the Fourier basis for obtaining the estimated smooth functions. In line with the work \cite{Liang2021116}, we applied the log-transformation on the non-negative precipitation data. Figure \ref{example_depend} 
depicts the clustering results, where the four clusters are represented by four different colors. The upper panels visualize the clustering results by the colored curves. The lower panels visualize the clustering results on the Canada map, where the filled circles indicate the locations of the 35 weather stations, and their colors indicate the cluster membership.

\section{Curve Registration}\label{CurveRegistration}
Functional data may differ in terms of two types of variation: amplitude variation in curve height and phase variation in lateral displacements of curve features (e.g. peaks, points of inflection, and threshold crossings). In the presence of phase variation, the functional data are commonly called time-warped functional data. For time-warped functional data, cross-sectional mean functions are un-interpretable: if important features such as peak locations randomly vary from curve to curve, ignoring the differences in timing when taking a cross-sectional mean will distort these features. The technique of transforming the arguments of curves to align various salient features is called curve registration, curve alignment, or time warping. The curve registration problem can be formulated as identifying the time warping function $h_i(\cdot)$ for each sample function $y_i$ such that the de-warped data $\{y_i\circ h_i: i=1, \ldots, n\}$ exhibit amplitude variation only, and all salient features are aligned. When $d=1$, the warping functions are commonly assumed to be smooth and strictly increasing and are diffeomorphisms of $\mathscr{T}$ to itself.

The choice of whether or not to perform curve registration depends on the problem at hand. For example, amplitude variation could be the main focus, with phase variation being a nuisance to be removed once identified. On the other hand, phase variation could contain all the information of interest, in contexts where issues such as timing are more important than relative peak heights. Finally, the joint variation between amplitude and phase can be central issues in the analysis. It turns out, for example, that there is a simple relation between the strength of a pubertal growth spurt and its timing, namely, that early spurts are stronger and later spurts are weaker \cite{24780816}. Park and Ahn \cite{10.1111/biom.12546} categorized three types of curve registration for vector-valued functional data: (1) component-wise warping, where the $n$ sample functions of each component random function are separately aligned to the component-wise template/target curve ($p$ templates in total); (2) subject-wise warping, where the $p$ sample functions associated with a subject are separately aligned to the subject-wise template curve ($n$ templates in total); (3) universal warping, where the $n$ sample functions of each component random function are time-warped by one identical warping function to the component-wise template curve ($p$ templates in total). Note that phase variation cannot be identified if the data exhibit both amplitude and phase variation, and the extraction depends very much on prior knowledge about how each type of variation is generated.

In most studies of functional data, curve registration is a pre-processing step and is independent of the following statistical analysis; see, e.g., \cite{Hall2007799}, \cite{SLAETS20122360}, \cite{FU2019159} and \cite{9626620}. Regarding Tier 3 categorization, the methods reviewed in Sections \ref{VectorSpace}-\ref{Dependent} either take the pre-processing strategy or leave the phase variation issue unaddressed. Only the seven articles explained below explicitly deal with phase variation, by integrating curve registration into clustering.

\subsection{Clustering in Finite-Dimensional Space}
In Liu and Yang \cite{LIU20091361}, given $z_i=k$, the formulation of $y_i$ is $y_i(t)=\alpha_{1i}+\mu_k(t+\alpha_{2i})$, where the distribution of $(\alpha_{1i}, \alpha_{2i})$ is $\mathscr{N}(\pmb{0}, \mbox{diag}(\gamma_1, \gamma_2))$; that is, each sample function differs from their mean function by a random shift in time and a random shift in evaluation. Applying the cubic B-spline basis expansion on $\mu_k$: $\mu_k(t+\alpha_{2i})=\sum_{v=1}^{m}a_{kv}b_v(t+\alpha_{2i})$, the final approximation of $y_i(t)$ is $y_i(t)=\alpha_{1i}+\sum_{v=1}^{m}a_{kv}[b_v(t)+\alpha_{2i}D^1b_v(t)]$, where $b_v(t)+\alpha_{2i}D^1b_v(t)$ is the linear Taylor approximation of $b_v(t+\alpha_{2i})$ at point $t$. Within the model-based clustering framework, an EM-type algorithm was developed, where $\{z_i, \alpha_{1i}, \alpha_{2i}\}_{i=1}^n$ were all treated as hidden data. The random effect $\alpha_{1i}$ is for amplitude variation, while the random effect $\alpha_{2i}$ is for phase variation.

In Wu and Hitchcock \cite{WU2016121}, with $\mathscr{T}=[0, 1]$, each warping function $h_i$ is approximated by the linear interpolation of the cumulative sum over $\{\theta_{i1}, \ldots, \theta_{iq}\}$, which are generated from a Dirichlet distribution. The formulation of the sample function $y_i$ is $y_i(t)=\alpha_{1i}\sum_{v=1}^ma_{iv}b_v(h_i(t))+\alpha_{2i}$, where $\alpha_{1i}$ is a random stretching factor, and $\alpha_{2i}$ is a random shifting factor. The coefficient vectors $\{\pmb{a}_i\}_{i=1}^n$ have a Gaussian mixture model. The prior on $\{\alpha_{1i}\}_{i=1}^n$ is a Gaussian distribution, on $\{\alpha_{2i}\}_{i=1}^n$ is a uniform distribution, and on $\pmb{\pi}$ is another Dirichlet distribution. Then the unknown model parameters and the mixing proportion vector are estimated simultaneously. The random effects $\alpha_{1i}$ and $\alpha_{2i}$ are for amplitude variation, while the phase variation is modeled by the nonparametric time-warping function.

In Maire et al. \cite{MAIRE201727}, the functional data are either curves or images. Each cluster is represented by a template function, and the cluster members are random deformations of the template function: given $z_i=k$, $y_i=\alpha_{1i}\mu_k(h(t; \pmb{\alpha}_{2i}))=\alpha_{1i}\sum_{v=1}^ma_{kv}b_v(h(t; \pmb{\alpha}_{2i}))$. The prior on $z_i$ is a multinomial distribution, on $\alpha_{1i}$ is a gamma distribution, on $[\pmb{\alpha}_{2i}|z_i=k]$ is a zero-mean Gaussian distribution with covariance matrix $\pmb{\Gamma}_k$. An EM-type algorithm was developed, where $\{z_i, \alpha_{1i}, \pmb{\alpha}_{2i}\}_{i=1}^n$ were all treated as hidden data. The random effect $\alpha_{1i}$ is for amplitude variation, while the phase variation is modeled by the parametric time-warping function.

\subsection{Clustering in Infinite-Dimensional Space}
The approach in \cite{Shi2008267} was again taken by Zeng et al. \cite{Zeng2019.1607744}, where each subject is characterized by both a random function $y_i$ (with $p=2$) and a covariate vector $\pmb{c}_i$, with the data denoted by $\{(\tilde{y}_i(\underline{t}_i), \pmb{c}_i): i=1, \ldots, n\}$. The decomposition of the sample function $y_i$, with $z_i=k$, is $y_i(t)=\mu_k(h_{ik}(t))+x_{ik}(t)$, where $h_{ik}$ is the inverse of a warping function, $\mu_k$ is the fixed effect, and $x_{ik}$ is the random effect; each mean function $\mu_k$ is approximated by the cubic Hermite spline basis, and each $x_{ik}^j$ ($j=1, 2$) is modeled by a zero-mean Gaussian process with the Mat\'{e}rn kernel. The tabular data $\{\pmb{c}_i\}_{i=1}^n$ are fitted by a logistic regression model to estimate the cluster membership values: $\Pr(z_i=k)=\frac{\exp(\pmb{c}_i^T\pmb{\beta}_k)}{1+\sum_{v=1}^{K-1}\exp(\pmb{c}_i^T\pmb{\beta}_v)}$. The $\pmb{\beta}_k$'s are estimated jointly with the other unknown parameters in an EM-type algorithm. The phase variation is modeled by the parametric time-warping function.

The works \cite{SANGALLI20101219}, \cite{Kaziska2011cs} and \cite{Guo2021777} all take the equivalent relation strategy for phase variation. Sangalli et al. \cite{SANGALLI20101219} defined the similarity between two functions as the cosine similarity between their derivatives: $\mathbbm{s}(y, \eta) = \frac{1}{p}\sum_{j=1}^{p} \frac{\langle D^1y^j, D^1\eta^j\rangle} {\|D^1y^j\| \|D^1\eta^j\|}$. Hence, the similarity measure is invariant to any warping function $h\in H=\{\alpha_1t+\alpha_2: \alpha_1\geq0\}$; that is, $\mathbbm{s}(y, \eta)=\mathbbm{s}(y\circ h, \eta\circ h)$ for any $h\in H$. A modified $k$-means algorithm was developed, where a cluster centroid $\mu_k$ was defined as $\mu_k=\sup_{\mu}\sum_{z_i=k}\sup_{h_i\in H}\mathbbm{s}(\mu, y_i\circ h_i)$, and a sample function $y_i$ was assigned to the cluster that has the largest value of $\sup_{h\in H}\mathbbm{s}(\mu_k, y_i\circ h)$.

In Kaziska \cite{Kaziska2011cs}, each parametric planar curve $y$,  with $d=1$ and $p=2$, is represented by its velocity functions $(\psi, \theta)$ in log-polar coordinates: $D^1y(t)=\exp(\psi(t))\exp(\ddot{\iota}\theta(t))$, where $\ddot{\iota}$ is the imaginary unit, $\psi$ is a log-speed function, and $\theta$ is the angle the velocity vector makes with a horizontal axis. $H$ is the set of all orientation-preserving diffeomorphisms from $\mathscr{T}$ to $\mathscr{T}$. For any $h\in H$, the reparameterized function $y\circ h$ is then represented by $(\psi, \theta)\ast h:=(\psi\circ h+\log(D^1h), \theta\circ h)$. Let $\mathscr{W}$ denote a pre-function space, which is an infinite-dimensional manifold. The group $H$ acts on $\mathscr{W}$ from the right, and hence $\mathscr{W}/\sim_H$ is a quotient space (not a manifold). The geodesic distance between any two functions $(\psi_i, \theta_i)$ and $(\psi_j, \theta_j)$ in $\mathscr{W}/\sim_H$ is $\mathbbm{d}_s((\psi_i, \theta_i), (\psi_j, \theta_j))=\min_{h\in H}\mathbbm{d}_c((\psi_i, \theta_i), (\psi_j, \theta_j)\ast h)$, where $\mathbbm{d}_s$ and $\mathbbm{d}_c$ are the geodesic distances between points on $\mathscr{W}/\sim_H$ and $\mathscr{W}$, respectively. With the constraint that planar curves that differ by orientation-preserving reparameterizations are to be viewed as representing the same shape, the distance between two planar curves $y_i$ and $y_j$, represented by $(\psi_i, \theta_i)$ and $(\psi_j, \theta_j)$, is defined as $\mathbbm{d}(y_i, y_j)=\mathbbm{d}_s((\psi_i, \theta_i), (\psi_j, \theta_j))$. The clustering objective is to minimize $\sum_{k=1}^K\frac{2}{n_k}\sum_{z_i=k, z_j=k}\mathbbm{d}(y_i, y_j)^2$, where $n_k$ is the size of the $k$th cluster.

The subspace clustering method developed by Guo et al. \cite{Guo2021777} is built on the square root velocity function (SRVF) representation \cite{5601739} (to be explained in the following section). The SRVF representation of any absolutely continuous and differentiable function $\eta$ is a square-integrable function $y:=\mbox{SRVF}(\eta)\in L^2 (\mathscr{T}, \mathbb{R}^p)$. $H$ is the set of all diffeomorphisms from $\mathscr{T}$ to $\mathscr{T}$, which is a Lie group that acts on $L^2 (\mathscr{T}, \mathbb{R}^p)$ from the right by composition. Then all orbits $[y]=\{y\circ h, \forall h\in H\}$ together define the quotient manifold $L^2 (\mathscr{T}, \mathbb{R}^p)/\sim_H$. They further introduced a rotation group and generalized the equivalence relation to include rotation; that is, two functions $y_i$ and $y_j$ are equivalent if they are reparameterized and rotated version of each other. The final set of unique equivalent classes is again a quotient manifold. Within the framework of subspace clustering, the original subspaces become subspaces in the quotient manifold. The self-expressive equation is in the form of $\sum_{j=1}^na_{ij}\mathscr{P}_{[y_i]}([y_j])=0$, where $[y_i]$ is the equivalent class that includes $y_i$, and $\mathscr{P}_{[y_i]}([y_j])$ is the projection of $[y_j]$ onto the tangent space rooted at $[y_i]$. A graph partitioning algorithm was applied on the affinity matrix $\pmb{A}=[a_{ij}]_{n\times n}$.

\begin{figure}[!ht]
	\centering
	\includegraphics[width=14cm]{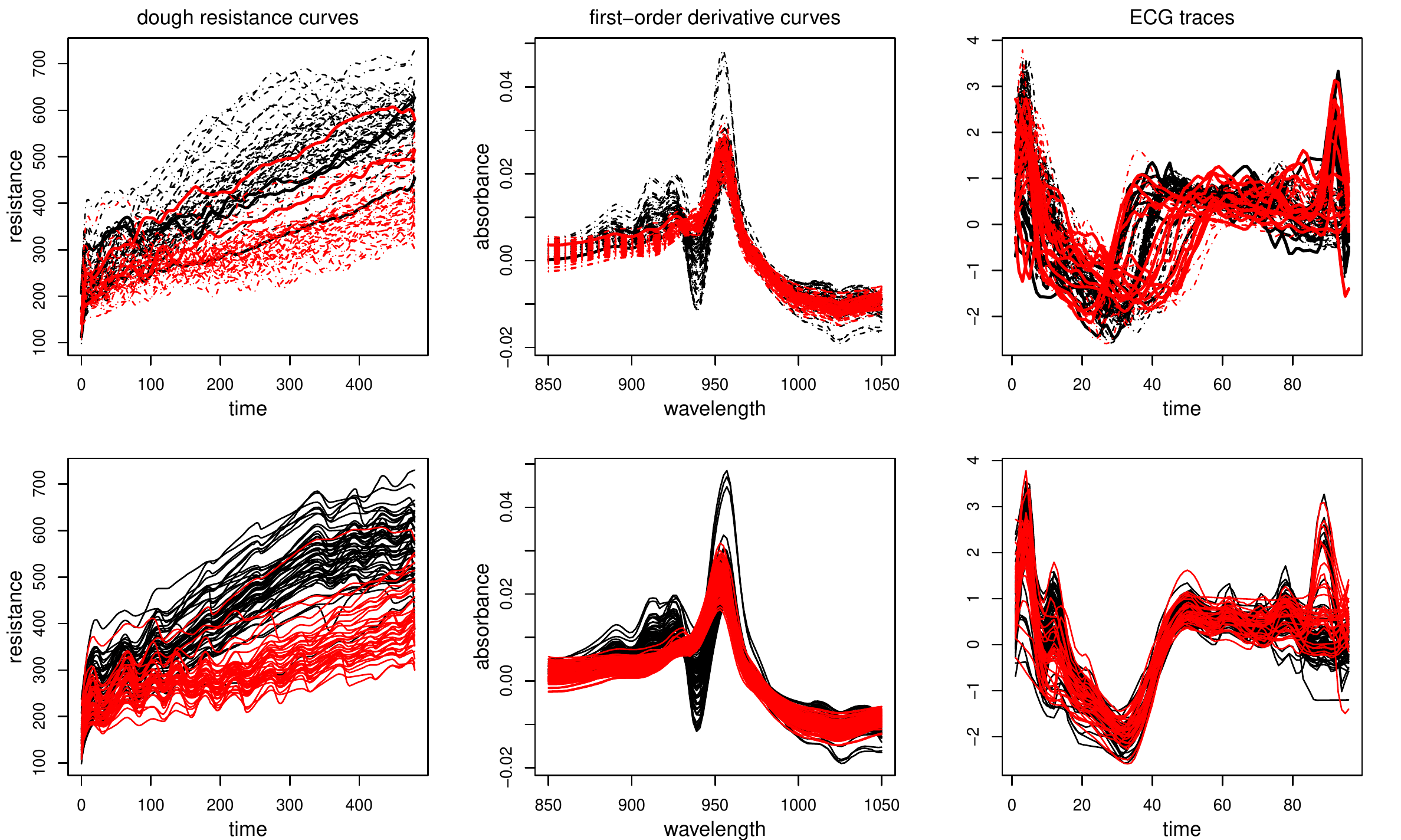}
	\caption{The upper panels demonstrate the clustering results, where a solid line means that the curve is incorrectly clustered. For example, a red solid line in the top left panel means that a bad-quality flour is incorrectly assigned to the good-quality cluster. For the Tecator data, only two derivative curves are incorrectly clustered, corresponding to one small-fat-percentage sample and one large-fat-percentage sample. However, for the ECG data, the clustering results are poor. The lower panels depict the aligned curves, where each curve is aligned to the Karcher mean function of its cluster. It is clear from the lower panels that salient features of the curves within a cluster are aligned. For the Tecator dataset, the functional data do not have phase variation, and therefore the curves before and after alignment are virtually identical.}
\label{example_align}
\end{figure}

In Figure \ref{example_align}, 
we apply the ``kmeans\_align'' algorithm in the R package \textit{fdasrvf} on three real datasets. The clustering method is built on the algorithm procedure in Sangalli et al. \cite{SANGALLI20101219} but replaces the similarity measure $\mathbbm{s}(y, \eta)$ with the Fisher-Rao distance metric \cite{arxiv.1103.3817}. The Tecator data are available at \url{http://lib.stat.cmu.edu/datasets/tecator}. The original data are near infrared absorbance spectra of 215 meat samples, among which 138 meat samples have a fat percentage of less than 20\%. Each sample was probed with a 100-channel absorbance spectrum within a wavelength range of 850–1050 nm. The curves in Figure \ref{example_align} are the first-order derivatives of the estimated smooth functions. The ECG dataset is taken from the UEA \& UCR Time Series Classification Repository (i.e., the ``ECG200'' dataset). The curves in Figure \ref{example_align} are from the training dataset, which consists of 69 time series from one class (in black) and 31 time series from another class (in red), all sampled at 96 time instants. The upper panels show the clustering results, where a curve in the form of bold solid line means that it is incorrectly clustered. The lower panels show the aligned curves, where each curve is aligned to the Karcher mean function of its cluster.  The algorithm procedure iterates between the clustering step and the curve registration step, and the final clustering results in the upper panels are obtained by applying the functional $k$-means algorithm on the aligned curves in the lower panels.

\section{Discussion}\label{NewFramework}
\subsection{A Synergistic Clustering Framework}
The above overview reveals that most documented works ignore the synergy between the smoothing step, the feature-extraction step (if any), and the clustering step. The upper line of approach (steps 1-3 in Figure \ref{twostep}) provides no assurance that the proxy extracted in steps 1 \& 2 is optimal for the subsequent cluster analysis in step 3, because they are implemented sequentially by optimizing different objective functions. Dimension reduction typically aims to retain as much variance as possible in as few dimensions as possible, whereas cluster analysis aims to find similar and dissimilar observations in the dataset and allocate the observations accordingly to clusters. Likewise, in the bottom line of approach, the smoothing step (step 1) is independent of the clustering step (step 2*) in most existing functional data clustering methods. However, smoothing can be integrated into the clustering algorithm. For example, one can jointly optimize the hyper-parameters in the smoothing step (e.g., the bandwidth parameter in a kernel smoother) and the clustering step. Moreover, when phase variation is a nuisance, then curve registration should be performed cluster-wise, instead of in the independent smoothing step.

Very few attempts have been made at integrating smoothing and clustering that reinforces the relationships between these two tasks. The objective function in Yamamoto \cite{Yamamoto2012219} is $\sum_{k=1}^K \sum_{z_i=k}\|y_i-\mathscr{P}\bar{y}_k\|^2=\sum_{i=1}^n \|y_i-\mathscr{P}y_i\|^2+\sum_{k=1}^K \sum_{z_i=k}\|\mathscr{P}y_i-\mathscr{P}\bar{y}_k\|^2$, where $\bar{y}_k$ is the sample average of the $y_i$'s with $z_i=k$, and $\mathscr{P}$ is the projection operator that projects a function onto the linear space spanned by $m$ orthonormal basis functions $\mathscr{B}=\{b_1 , \ldots, b_m\}$. The algorithm minimizes the objective function over the variables $\{z_i: i=1, \ldots, n\}$ and the basis functions $\{b_1, \ldots, b_m\}$ iteratively until convergence. Yamamoto and Terada \cite{YAMAMOTO2014133} modified the objective function to keep only the within-cluster variation in the projected space: $\min_{\mathscr{P}, \{z_i\}_{i=1}^n}\sum_{k=1}^K \sum_{z_i=k}\|\mathscr{P}y_i-\mathscr{P}\bar{y}_k\|^2$. Yamamoto and Hwang \cite{Yamamoto2017294} assumed that the subspace optimal for clustering is spanned by only a subset of basis functions from $\mathscr{B}$, denoted by $\mathscr{B}_c$, and introduced a penalty into the objective function:
\begin{align*}
\min_{\mathscr{P}, \mathscr{P}_c, \{z_i\}_{i=1}^n} \sum_{i=1}^n \|y_i-\mathscr{P}y_i\|^2+\theta_1\sum_{k=1}^K \sum_{z_i=k}\|\mathscr{P}_cy_i-\mathscr{P}_c\bar{y}_k\|^2 + \theta_2\sum_{k=1}^K n_k\|\mathscr{P}_c\bar{y}_k\|^2,
\end{align*}
where $n_k$ is the size of the $k$th cluster, and $\mathscr{P}_c$ is the projection operator that projects a function onto the linear space spanned by the basis functions in $\mathscr{B}_c$.

\begin{figure}[!ht]
	\centering
	\includegraphics[width=12cm]{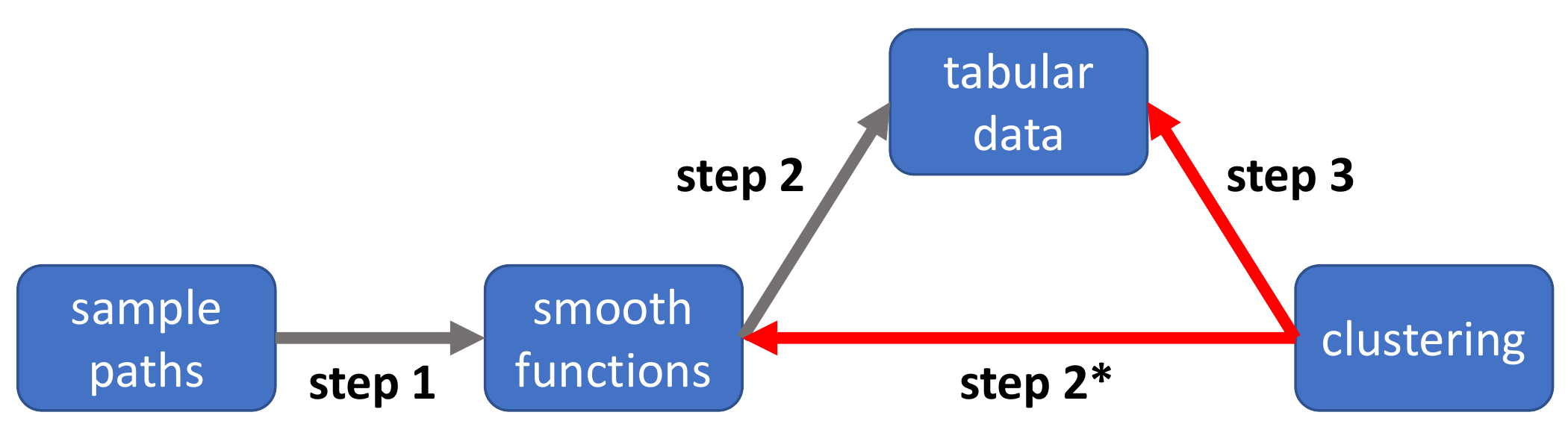}
	\caption{A new methodological framework: in the upper line of approach, we perform joint smoothing, feature learning, and clustering; in the bottom line of approach, we perform joint smoothing and clustering. We might call this new framework as the synergistic clustering framework.}
\label{twostepnew}
\end{figure}

The documented literature is focused on creating proxies for sample paths, either estimated smooth functions or extracted tabular data, and then applying traditional clustering algorithms on the proxies; there seems to be either no or insufficient justification why a particular multivariate clustering algorithm was chosen. Motivated by the deficiency in the current tandem approach, we here propose a new methodological framework for functional data clustering, as summarized by Figure \ref{twostepnew}.
The difference between Figure \ref{twostep} and Figure \ref{twostepnew} is that the directions of step 2* and step 3 are reversed, implying that step 1 and step 2* (or, steps 1 \& 2 and step 3) will coordinate with each other to obtain the true clusters for the original sample paths, not for any type of proxy. The problem with the current practice is that, in step 1 of constructing the functional form, different choices of functional basis will produce different groupings, even with the same clustering method. Therefore, to reveal the true grouping pattern underlying the sample paths, one should integrate step 1 and step 2* (or, steps 1 \& 2 and step 3). The principles behind our methodological framework are not new. In fact, the deficiency of the tandem approach has been documented in many works in the field of dimensionality reduction. For example, Aljalbout et al. \cite{ElieAljalbout2018} pointed out the same problem with most deep representation-learning based clustering methods, and proposed a new method that jointly optimizes the feature-extraction step and the clustering step.

\subsection{Shape Analysis}
Shape analysis makes use of sophisticated tools from differential and Riemannian geometry and group theory and is of fundamental importance to the cluster analysis of functional data \cite{Bauer2014, A.Srivastava2016functional}. Recall that, in the curve registration problem, two functions $x, y\in L^2 (\mathscr{T}, \mathbb{R})$ are considered equivalent if they can be time-warped to each other. The curve registration problem can naturally be posed as a shape analysis problem, where two functions are considered equivalent if there exists a (nice) homeomorphism $h: \mathscr{T}\mapsto\mathscr{T}$ such that $y=x\circ h$; that is, $x$ and $y$ differ only by composition with a homeomorphism. In addition to curves,  differential-geometric tools are particularly powerful for the shape analysis of images and surfaces. We here give a brief explanation of the problem through the equivalence relation of re-parameterization. Other desired types of invariance are rotation, translation, and scaling. By directing the interest of the readers to this ambitious research field, we hope to motivate more efforts on  developing clustering methods that are invariant to shape-preserving transformations.

Let $(\mathscr{W}, \tau)$ be a topological space, where $\mathscr{W}$ is a set, and $\tau$ is a collection of open subsets of $\mathscr{W}$. For example, the Hilbert space $L^2 (\mathscr{T}, \mathbb{R}^p)$ with the $L^2$ inner product is a topological space. Let $H=\{h: \mathscr{T}\mapsto\mathscr{T}\}$ denote a set of homeomorphisms, where each $h$ is a bijection, and both $h$ and its inverse $h^{-1}$ are continuous. Note that, $H$ need be a group under composition. The group $H$ defines the equivalence relation on $\mathscr{W}$, denoted by $\sim_{H}$, through the action of $H$ on $\mathscr{W}$ from the right by composition: $x\sim_{H}y$ if and only if there exists a homeomorphism $h\in H$ such that $y=x\circ h$. The equivalent class (a.k.a. orbit) that includes $w\in\mathscr{W}$, denoted by $[w]$, is the set $\{w\circ h: h\in H\}$. The equivalence relation $\sim_{H}$ on $\mathscr{W}$ determines a new set, called the quotient set and denoted by $\mathscr{W}/\sim_{H}$, whose elements are the distinct equivalence classes; that is, an equivalence relation defines a partition of $\mathscr{W}$ by the equivalence classes. The quotient map associated with $\sim_{H}$ refers to the (surjective) map from $\mathscr{W}$ to $\mathscr{W}/\sim_{H}$: $q(w)=[w]$. For any subset $\mathscr{S}\subseteq\mathscr{W}/\sim_{H}$, the following holds: $q^{-1}(\mathscr{S})=\{w\in\mathscr{W}: [w]\in \mathscr{S}\}$. The quotient space under $\sim_{H}$ is the quotient set $\mathscr{W}/\sim_{H}$ equipped with the quotient topology, which is the topology whose open sets are the subsets $\mathscr{S}\subseteq\mathscr{W}/\sim_{H}$ such that $q^{-1}(\mathscr{S})$ is an open subset of $\mathscr{W}$; that is, $\mathscr{S}\subseteq\mathscr{W}/\sim_{H}$ is open in the quotient topology on $\mathscr{W}/\sim_{H}$ if and only if $q^{-1}(\mathscr{S})\in\tau$.
The goal is to make the quotient set $\mathscr{W}/\sim_{H}$ into a metric space such that we can readily calculate the distance between two equivalent classes. The importance of having well-defined shape spaces with metrics and geodesics is that they are essential for statistical analysis. In particular, given shape metrics, sample statistics, and probability models on shape spaces, it will become possible to extend finite-dimensional clustering methods to shape spaces. There are two major challenges though: (1) getting an appropriate metric $\mathbbm{d}$ on the quotient set $\mathscr{W}/\sim_{H}$, and (2) developing an efficient algorithm for calculating $\mathbbm{d}([x], [y])$, for any $x, y\in\mathscr{W}$.

We here explain the popular SRVF representation. In Srivastava et al. \cite{5601739} and other related papers, $\mathscr{W}$ is the set of absolutely continuous functions $y: [0, 1]\mapsto\mathbb{R}^p$ and $y(0)=0$. The set $H$ is a Lie group of diffeomorphisms: $h$ is absolutely continuous, $h(0)=0$, $h(1)=1$, and $D^1h(t)>0$ almost everywhere. They introduced a bijection $\psi$ between $\mathscr{W}$ and $L^2 ([0, 1], \mathbb{R}^p)$, i.e., the SRVF representation, which changes the problem to defining an action of $H$ on $L^2 ([0, 1], \mathbb{R}^p)$ and then a metric on the quotient set $L^2 ([0, 1], \mathbb{R}^p)/\sim_{H}$. In particular, the action of $H$ on $\mathscr{W}$ from the right by composition translates to an action of $H$ on $L^2 ([0, 1], \mathbb{R}^p)$ given by $y\ast h(t)=y(h(t))\sqrt{D^1h(t)}$. More importantly, the action is by isometries: $\langle x, y\rangle=\langle x\ast h, y\ast h\rangle$, $\forall h\in H$. Define a function $\mathbbm{d}: L^2 ([0, 1], \mathbb{R}^p)/\sim_{H}\times L^2 ([0, 1], \mathbb{R}^p)/\sim_{H}\mapsto\mathbb{R}$ by $\mathbbm{d}([x], [y])=\inf_{h_1, h_2\in H}\|x\ast h_1-y\ast h_2\|=\inf_{h\in H}\|x-y\ast h\|$, where the second equality follows from the fact that $H$ acts by isometries. If each equivalent class $[x]$ is closed in $L^2 ([0, 1], \mathbb{R}^p)$, then $\mathbbm{d}$ is a metric on the quotient space $L^2 ([0, 1], \mathbb{R}^p)/\sim_{H}$. Moreover, the quotient topology coincides with the topology induced by the metric $\mathbbm{d}$.

\subsection{Other Topics}
In the most common setting for functional data analysis, the basic unit of observation is the real-valued curve. The articles reviewed above all take up the setting that $Y(t)$ is a real-valued random variable/vector. In longitudinal data analysis, one frequently encounters non-continuous data that are repeatedly collected for a sample of individuals over time. The repeated observations could be binomial, Poisson or of another discrete type. Hall et al. \cite{WOS:000257674700004} extended the fPC technique of \cite{doi:00001745} to non-continuous functional data by positing a latent continuous stochastic process that, through a known link function, gives rise to the observed non-continuous outcome. In contrast to the rich literature for real-valued functional data, very few works exist for clustering non-continuous functional data. The idea of latent factor in Hall et al.  \cite{WOS:000257674700004} was adopted by Huang et al. \cite{HuangLiGuan} and Lim et al. \cite{Lim20203205} for clustering non-continuous functional data. In Huang et al. \cite{HuangLiGuan}, each $Y(t)$ is a binary random variable. They assumed that the latent factor is a Gaussian process and that, conditioned on the latent factor, the distribution of $Y(t)$ belongs to the canonical exponential family. Lim et al.  \cite{Lim20203205} proposed a model-based clustering method for discrete functional data with $p\geq2$. They also assumed that there exists a latent Gaussian process for each component random function $Y^j$. In Kom\'{a}rek and Kom\'{a}rkov\'{a} \cite{Komarek2013177}, the random vector $Y(t)=(Y^1(t), \ldots, Y^p(t))^T$ is composed of both continuous and discrete random variables; the motivating example is a longitudinal study on patients with primary biliary cirrhosis with a continuous bilirubin level, a discrete platelet count, and a dichotomous indication of blood vessel malformations. They adopted the mixed-effects model with a Gaussian mixture in the distribution of random-effect coefficient vectors.

Table \ref{taxonomygroup} in the appendix clearly shows that the literature is dominated by a few categories: the model-based and centroid-based clustering categories for extracted tabular data, and the centroid-based and (new)dissimilarity categories for estimated smooth functions. To date, very few functional data clustering studies made use of more recently developed clustering algorithms, e.g., density-based clustering methods \cite{JT2021ICDM}. The field could be greatly advanced if state-of-the-art multivariate clustering methods are carefully customized to functional data in that smoothing and curve registration are properly integrated.

Most clustering methods for univariate functional data are incapacitated for vector-valued functional data, mainly due to the complex dependency among the component random functions. To avail of modern clustering methods, we need a powerful feature-learning tool such that the extracted finite features maximally explain the variation in the vector-valued functional data. A research direction is to apply deep representation learning methods for non-linear dimensionality reduction and perform clustering in the lower-dimensional latent space. For example, for moving-object-trajectory clustering, Yao et al. \cite{WOS:000426968704018} applied the seq2seq LSTM model (a encoder-decoder architecture built on long short-term memory networks) to obtain a fixed-length vector representation of each (variable-length) feature-vector sequence, where the feature-vector sequences were extracted from the raw trajectories. However, the seq2seq LSTM model was trained by minimizing the reconstruction error, ignoring completely the clustering objective.

Another pressing issue in functional data clustering is the scalability of a clustering method (including the smoothing step and, if any, the feature-learning step). It is desirable to develop a highly scalable clustering algorithm, and evaluate the performance of an algorithm in terms of accuracy, efficiency, and scalability. Therefore, in addition to an efficient algorithmic design, an efficient implementation technique is needed. For example, different tree-based data structures, such as k-dimensional tree and R-tree, have been implemented to improve the performance of clustering big data \cite{3416921.3416942}.

Finally, unlike other data-mining areas for which various open-source software and data repositories are available, little effort has been made to foster open functional data analysis, in terms of both software and data. This poses a challenge to compare existing clustering algorithms and evaluate new algorithms. It is urgent to have an open functional-data repository, containing datasets from different fields of application, to facilitate the evaluation and comparison of clustering models.

\section{Conclusion}
In this paper, we surveyed all methodological studies on the subject of functional data clustering. These studies can be structured according to a three-tier categorization, depending upon whether they work directly with the estimated smooth functions or indirectly with extracted tabular data (Tier 1), the type of the clustering algorithm (Tier 2), and how they address phase variation and/or amplitude variation (Tier 3). For the cross-pollination of ideas across germane research fields, we also included a few good references on the cluster analysis of time series, trajectory data and spatio-temporal data. The limitation of past studies and some potential topics for future study were also discussed.

\section*{Acknowledgement}
This publication has emanated from research supported by a research grant from Science Foundation Ireland (SFI) under grant number 16/RC/3872 and is co-funded under the European Regional Development Fund.

\section*{Appendix}
\appendix
Articles included in this review are selected in the following manner. Search for the literature on functional data clustering was carried out in two databases: Scopus and Web of Science.
\begin{enumerate}
  \item Any article that contains the term ``functional data'' and any word starting with ``cluster'' (i.e., cluster, clusters, clustered, clustering) in either the title, abstract or keyword list was treated as potentially relevant. Additionally, any article that contains the term ``functional data'' and the word ``unsupervised'' in either the title, abstract or keyword list was also selected.
  \item Non-English articles and non-peer reviewed articles were excluded. Duplicate articles between the two databases were then removed.
  \item The term "functional data" is also used in the fields of genetics and molecular biology, where the word function means functionality. All such irrelevant articles were excluded.
\end{enumerate}
The cleaning process gave rise to 367 articles that are relevant to the cluster analysis of functional data.
\begin{enumerate}
\setcounter{enumi}{3}
 \item We only review methodological articles. Application-only articles were excluded.
\end{enumerate}
A total of 103 articles ended up meeting the inclusion criteria. Five articles directly deal with the (high-dimensional) vectors of function evaluations, and were also excluded from our review, due to their limited applicability and low efficiency. The five survey articles are separately explained in Section \ref{Intro}. The remaining 93 articles contribute new functional data clustering methods and are grouped in Table \ref{taxonomygroup} according to our taxonomy.
\begin{table}[!ht]
  \centering
  \caption{Tier 1 and Tier 2 categorizations of existing functional data clustering methods.}\label{taxonomygroup}
  \begin{tabular}{lll}
    \hline
    \hspace{-0.5cm}Tier 1& Tier 2& Refs\\
    \hline
     \parbox[c]{0mm}{\multirow{6}{*}{\rotatebox[origin=c]{90}{Vector Space}}} & centroid-based clustering&  \cite{Abraham2003} \cite{Antoniadis2013} \cite{Delaigle2019271} \cite{10.1111/rssc.12404} \cite{LUZLOPEZGARCIA2015231} \cite{Garcia-Escudero2005185} \cite{journal.pone.0242197} \cite{KIM2020104626} \cite{Tarpey200734} \cite{Tarpey2003}\\
        & ensemble clustering& \cite{Abramowicz2019} \cite{WOS:000408988700020} \cite{SECCHI201353} \\
        & model-based clustering&  \cite{BENSLIMEN201897} \cite{BenSlimen20} \cite{10.1214/15-AOAS861} \cite{Bouveyron20151143} \cite{Bouveyron2011281} \cite{bouveyron:hal-02862177} \cite{Giacofci201331} \cite{GOLOVKINE2022107376} \cite{JACQUES2013164} \cite{JACQUES201492} \cite{JamesSugar} \cite{Komarek2013177} \cite{Liang2021116} \cite{LIU20091361} \\
        & & \cite{Ma2008625} \cite{MAIRE201727}  \cite{Nguyen201676} \cite{Nguyen20185} \cite{Rivera-Garca2019201} \cite{Schmutz20201101}  \cite{Sharp2021735} \cite{9679476} \cite{WuWang2021} \cite{WU2016121}\\
        & nonparametric Bayesian& \cite{Margaritella2021167}  \cite{RayMallick} \cite{White2021586}\\
        & miscellaneous& \cite{Kayano2010211} (self-organizing map) \cite{Serban2012805} (multi-level clustering)\\
    \hline
    \parbox[c]{0mm}{\multirow{10}{*}{\rotatebox[origin=c]{90}{Function Space}}} & centroid-based clustering&  \cite{CUESTAALBERTOS20074864} \cite{HEBRAIL20101125} \cite{Ieva2013401} \cite{LALOE202151}  \cite{Meng2018166}  \cite{Romano2017645} \cite{Tokushige20071} \cite{Yamamoto2012219} \cite{Yamamoto2017294} \cite{YAMAMOTO2014133} \cite{Zambom2019527} \cite{Zambom2022}\\
        & covariance-based clustering& \cite{Ieva20161} \cite{Kashlak2019214} \\
        & hierarchical clustering& \cite{Chen2021425} \cite{DABONIANG20074878} \cite{FerreiraHitchcock2009} \cite{Lim2019368} \\
        & new (dis)similarity&  \cite{Bruno2011975} \cite{biom.12161Chen} \cite{FLORIELLO20171} \cite{Gaetan2016964} \cite{Giraldo2012403} \cite{Haggarty2015491}  \cite{Kaziska2011cs} \cite{Li201615}  \cite{Martino2019301} \cite{Romano2020} \cite{SANGALLI20101219} \cite{10.1214/14-BA925} \\
        & &\cite{Tzeng20183492} \cite{vanDelft2021469}\\
        & nonparametric Bayesian&  \cite{Nguyen2010817} \cite{Nguyen20111249} \cite{Petrone2009755}\\
        & regression mixture& \cite{Chamroukhi2016374} \cite{HuangLiGuan} \cite{cjs.11680} \cite{Lim20203205} \cite{Shi2008267} \cite{Zeng2019.1607744} \cite{Zhong2021852}\\
        & subspace clustering& \cite{pmlr-v37-bahadori15} \cite{Chiou2007} \cite{ChiouLi20082090} \cite{Guo2021777} \cite{10.1111/biom.12546} \\
        & miscellaneous& \cite{Baragilly202247} (forward search) \cite{doi2018.1494280} (self-organizing map)\\
        & & \cite{Ciollaro20162922} (density-based clustering) \cite{Galvani2021} (functional Cheng and Church) \\
    \hline
  \end{tabular}
\end{table}

\bibliographystyle{unsrt}
\bibliography{Ref_FCreview}
\end{document}